\documentclass[sigconf]{acmart}

\usepackage{enumitem}
\usepackage{algorithm}
\usepackage{algpseudocode}
\usepackage{multicol}
\usepackage{caption}
\usepackage{subcaption}
\usepackage{booktabs} 
\usepackage{siunitx} 
\usepackage{etoolbox}
\usepackage{soul}
\usepackage{cancel}
\usepackage[normalem]{ulem}

\usepackage[font=small,skip=0pt]{caption}

\usepackage{amsmath,algorithm,tabularx}
\usepackage{algpseudocode}

\makeatletter
\newcommand{\multiline}[1]{%
  \begin{tabularx}{\dimexpr\linewidth-\ALG@thistlm}[t]{@{}X@{}}
    #1
  \end{tabularx}
}

\makeatother
\AtBeginDocument{%
  \providecommand\BibTeX{{%
    \normalfont B\kern-0.5em{\scshape i\kern-0.25em b}\kern-0.8em\TeX}}}

\setcopyright{acmcopyright}

\copyrightyear{2023}
\acmYear{2023}
\setcopyright{rightsretained}
\acmConference[ICS '23]{2023 International Conference on
Supercomputing}{June 21--23, 2023}{Orlando, FL, USA}
\acmBooktitle{2023 International Conference on Supercomputing (ICS '23), June
21--23, 2023, Orlando, FL, USA}
\acmDOI{10.1145/3577193.3593715}
\acmISBN{979-8-4007-0056-9/23/06}

\begin{document}

\pdfcompresslevel=9
\pdfminorversion=5
\pdfobjcompresslevel=2



\newcommand{\yujia}[1]{\textcolor{red}{#1}} 
\newcommand{\shixun}[1]{\textcolor{blue}{#1}} 

\title{Anatomy of High-Performance GEMM with Online Fault Tolerance on GPUs}

\author{Shixun Wu}
\affiliation{%
  \institution{University of California, Riverside}
  \city{Riverside}
  \state{CA}
  \country{USA}}
\email{swu264@ucr.edu}
\authornote{Shixun Wu and Yujia Zhai contributed equally to this paper.}

\author{Yujia Zhai}
\affiliation{%
  \institution{University of California, Riverside}
  \city{Riverside}
  \state{CA}
  \country{USA}}
\email{yzhai015@ucr.edu}
\authornotemark[1]

\author{Jinyang Liu}
\affiliation{%
  \institution{University of California, Riverside}
  \city{Riverside}
  \state{CA}
  \country{USA}}
\email{jliu447@ucr.edu}

\author{Jiajun Huang}
\affiliation{%
  \institution{University of California, Riverside}
  \city{Riverside}
  \state{CA}
  \country{USA}}
\email{jhuan380@ucr.edu}

\author{Zizhe Jian}
\affiliation{%
  \institution{University of California, Riverside}
  \city{Riverside}
  \state{CA}
  \country{USA}}
\email{zjian106@ucr.edu}

\author{Bryan M. Wong}
\affiliation{%
  \institution{University of California, Riverside}
  \city{Riverside}
  \state{CA}
  \country{USA}}
\email{bryan.wong@ucr.edu}

\author{Zizhong Chen}
\affiliation{%
  \institution{University of California, Riverside}
  \city{Riverside}
  \state{CA}
  \country{USA}}
\email{chen@cs.ucr.edu}

\begin{abstract}

  General Matrix Multiplication (GEMM) is a crucial algorithm for various applications such as machine learning and scientific computing since an efficient GEMM implementation is essential for the performance of these calculations. While researchers often strive for faster performance by using large computing platforms, the increased scale of these systems can raise concerns about hardware and software reliability. In this paper, we present a design of a high-performance GPU-based GEMM that integrates an algorithm-based fault tolerance scheme that detects and corrects silent data corruptions at computing units on-the-fly. We explore fault-tolerant designs for GEMM at the thread, warp, and threadblock levels, and also provide a baseline GEMM implementation that is competitive with or faster than the state-of-the-art, closed-source cuBLAS GEMM. We present a kernel fusion strategy to overlap and mitigate the memory latency due to fault tolerance with the original GEMM computation. To support a wide range of input matrix shapes and reduce development costs, we present a template-based approach for automatic code generation for both fault-tolerant and non-fault-tolerant GEMM implementations. We evaluate our work on NVIDIA Tesla T4 and A100 server GPUs. Our experimental results demonstrate that our baseline GEMM shows comparable or superior performance compared to the closed-source cuBLAS. Compared with the prior state-of-the-art non-fused fault-tolerant GEMM, our optimal fused strategy achieves a 39.04\% speedup on average. In addition, our fault-tolerant GEMM incurs only a minimal overhead (8.89\% on average) compared to cuBLAS even with hundreds of errors injected per minute. For irregularly shaped inputs, the code generator-generated kernels show remarkable speedups of $160\% \sim 183.5\%$ and $148.55\% \sim 165.12\%$ for fault-tolerant and non-fault-tolerant GEMMs, respectively, which outperforms cuBLAS by up to $41.40\%$.
  
  
\end{abstract}




\settopmatter{printfolios=true}


\maketitle
\section{Introduction}

The growing complexity and scale of modern computing systems have made them more vulnerable to transient faults, which are errors that can occur during the transfer of signals or storage of values. They can be caused by various factors such as shrinking transistor width, higher circuit density, and lower near-threshold voltage operations \cite{laprie1985dependable, lutz1993analyzing, nicolaidis1999time}. These faults can have a significant impact on the reliability of computing systems, as demonstrated by the numerous instances of transient faults causing server crashes or rendering entire clusters of computers unusable in the past \cite{gomez2015detecting, li2012classifying}. In fact, the U.S. Department of Energy has identified reliability as one of the major challenges for exascale computing, highlighting the importance of addressing this issue \cite{lucas2014doe}.

Transient faults can threaten the reliability of even the most advanced computing systems, as demonstrated by simulations of an exascale machine with 190,000 cutting-edge Xeon Phi processors that could still experience daily transient errors even when protected by error correcting code \cite{oliveira2017experimental}. To protect software systems from computing errors at runtime and maintain their reliability, various techniques for fault tolerance have been proposed and studied by both academia and industry \cite{may1979alpha, baumann2002soft, geist2016supercomputing}.

There are two types of errors that can occur when a transient fault affects an application: fail-stop errors that cause the application to crash and fail-continue errors that allow the application to continue running but produce incorrect results. Fail-stop errors can often be mitigated through checkpoint/restart mechanisms \cite{phillips2005scalable, NEURIPS2019_9015, tao2018improving, tensorflow2015-whitepaper} or algorithmic approaches \cite{hakkarinen2014fail, chen2008scalable, chen2008extending}, but fail-continue errors can be more dangerous as they can corrupt application states without any warning and lead to incorrect computing results \cite{mitra2014resilience, cher2014understanding, dongarra2011international, calhoun2017towards, snir2014addressing}. These types of errors can be particularly problematic in safety-critical scenarios \cite{li2017understanding}. In this paper, we focus specifically on fail-continue errors from computing logic units (such as when a calculation produces an incorrect result) and assume that fail-stop and memory errors are protected by checkpoint/restart and error-correcting code, respectively. We refer to these types of errors as {\it soft errors}.

One approach to handling soft errors is through the use of dual modular redundancy (DMR). DMR involves duplicating computing instructions and inserting check instructions into the original program, typically with the assistance of compilers \cite{oh2002error, oh2002control, reis2005swift, yu2009esoftcheck, chen2016simd}. While DMR is a general approach that can be applied to any application, it can also introduce significant overhead, particularly for compute-intensive applications that require duplication of all computations. To reduce the overhead of fault tolerance, algorithm-based fault tolerance (ABFT) schemes have been developed for various applications in recent years. There have been numerous ABFT schemes developed for various applications in recent years. For example, Huang and Abraham proposed the first ABFT scheme for matrix-matrix multiplication \cite{huang1984algorithm}, Di and Cappello proposed an adaptive impact-driven fault tolerance approach for real-world HPC applications \cite{di2016adaptive}, and Chien at al. proposed the Global View Resilience system, a library that enables applications to efficiently add resilience \cite{chien2015versioned}. There are also ABFT schemes for widely-used algorithms such as sorting \cite{li2019ft}, fast Fourier transforms \cite{liang2017correcting, antola1992fault, tao1993novel}, iterative solvers \cite{chen2013online, tao2016new, chen2014extending}, and convolutional neural networks \cite{zhao2020algorithm}.

GEMM is a crucial building block for various applications in machine learning and scientific computing, which are often run on large computing facilities with accelerators such as GPUs for extended periods of time. As a result, the need for resilience against GEMM soft errors on GPUs is critical and has been extensively studied on both CPUs \cite{chen2008algorithm, gunnels2001fault, wu2013line} and GPUs \cite{ding2011matrix, kosaian2021arithmetic}. Existing works, which have demonstrated the feasibility to deploy ABFT for GEMMs on GPUs, either lack the architectural-aware designs and optimizations to mitigate the memory cost introduced by ABFT \cite{ding2011matrix} or are incapable of correcting the detected errors on-the-fly. To address these issues, we first build a high-performance GEMM kernel from scratch, which serves as the foundation for our novel and efficient approach that fuses the memory operations of ABFT with the original GEMM memory footprint through the implementation of custom fault-tolerant GPU kernels. We have also designed a template-based code generation scheme that eases kernel development efforts to support a wide range of input matrix shapes. More specifically, our contributions include the following:

\begin{itemize}
    \item We begin our work by optimizing the original GEMM. Through a series of tuning strategies including tiling, register reuse, and prefetching at both the register and shared memory levels, our GEMM achieves performance comparable to or faster than the closed-source cuBLAS GEMM. We provided a public GitHub repository to archive the related source codes of this paper \footnote{\href{https://github.com/shixun404/Fault-Tolerant-SGEMM-on-NVIDIA-GPUs.git}{\textcolor{blue}{\underline{https://github.com/shixun404/Fault-Tolerant-SGEMM-on-NVIDIA-GPUs.git}}}}.
    \item We detail the anatomy of the high-performance GEMM with online fault tolerance on GPUs from the thread level,  warp level, and  threadblock level.
    \item We propose a template-based code generation strategy to automatically generate high-performance GEMM kernels with or without fault tolerance for various input shapes.
    \item We benchmark our work on an NVIDIA Tesla T4 GPU and an NVIDIA A100 GPU. Compared with the prior state-of-art non-fused fault-tolerant baseline, our kernel fusion strategy shows a 39.04\% speedup on average while maintaining the online error correction functionality. The fault-tolerant version of the GEMM also shows a negligible overhead, 8.89\% on average, compared to the closed-source cuBLAS even when hundreds of errors are injected per minute. Additionally, for irregularly shaped inputs, the code generator-generated kernels achieve significant speedups of $160\% \sim 183.5\%$ and $148.55\% \sim 165.12\%$ for both fault-tolerant and non-fault-tolerant GEMMs, respectively, surpassing cuBLAS by a maximum of $41.40\%$.

\end{itemize}

The rest of the paper is organized as follows: we introduce the background and related works in Section \ref{sec:background} and  describe our designs and optimizations for the original GEMM and fault-tolerant GEMM in Sections \ref{sec:design_and_optimizations} and \ref{sec:fault_tolerant}. Evaluation results are given in Section \ref{sec:results}. We conclude our paper and present future work in Section \ref{sec:conclusion}.

\section{Background and Related Works}
\label{sec:background}

In this chapter, we provide an overview of two main categories of software approaches, duplication-based and algorithm-based approaches, to tolerate runtime computing errors. We also present prior works that employ the kernel fusion strategy to boost applications, especially those centered on GEMM operations.

\subsection{Duplication-Based Fault Tolerance}

Duplication-based fault tolerance, also known as Dual Modular Redundancy (DMR), has been a common protection scheme from software errors for many years. This technique, which is rooted in compiler-assisted approaches, has been extensively studied and is classified based on the Sphere of Replication (SoR), or the logical domain of redundant execution. Previous work in duplication-based fault tolerance can be categorized into three cases: (1) Thread Level Duplication (TLD), which duplicates the entire processor and memory system, (2) TLD with Error Correction Code assumption (TLD+ECC), which duplicates the instructions but only loads operands from the same memory address, and (3) DMR only for computing errors, which duplicates only the computing instructions to prevent faulty results from being written to memory.

\subsection{Algorithm-Based Fault Tolerance}

Soft error protection algorithms are designed to detect and correct errors that may occur during the computation of iterative or computing-intensive applications. These algorithms have a long history of success, with the first such algorithm being developed for matrix-matrix multiplication in 1984 \cite{huang1984algorithm}. The basic idea behind these algorithms is to encode matrices $A$ and $B$ into checksum forms $A^c$ and $B^r$, respectively, using the following equations: 

\begin{equation}
A\xrightarrow[]{encode}A^c:=\left[\begin{array}{c}A \\ e^T A\end{array}\right],
\label{eqn:A_encode}
\end{equation}

\begin{equation}
B\xrightarrow[]{encode}B^r:=\begin{bmatrix}B & Be\end{bmatrix},
\label{eqn:B_encode}
\end{equation}
where $e$ is a transposed identity vector, [$1,1,\dots,1$]$^T$. The encoded matrices are then multiplied together to produce a matrix $C^f$ that contains both the correct result and checksum information: 

\begin{equation}
C^f = A^c\cdot B^r = \begin{bmatrix}C & Ce\\ e^TC & \end{bmatrix} = \begin{bmatrix}C & C^r\\ C^c & \end{bmatrix}.
\label{eqn:C}
\end{equation}

The accuracy of the final result can be verified by comparing the values in matrix $C$ to the checksum values in $C^r$ and $C^c$. If the difference between these values exceeds a predetermined threshold, it indicates that an error occurred during the computation. The cost of encoding and verifying the matrices using checksums is generally much smaller than the cost of the matrix multiplication itself, making this a lightweight method for detecting errors. These checksum algorithms can be used with any matrix multiplication algorithm, and the accuracy of the final result can be verified either online (during the computation) or offline (after the computation is complete).
 
Chen et al. \cite{chen2008algorithm} proposed an outer-product matrix-matrix multiplication, where the checksum relationship can be maintained during the middle of the computation:

\begin{equation}
C^f = \sum_sA^c(:,s)\cdot B^r(s,:) = \sum_s\begin{bmatrix}C_s & C_se\\ e^TC_s & \end{bmatrix}.
\end{equation}
Here $s$ is the step size of the outer-product update on matrix $C$, and $C_s$ represents the result of each step of the outer-product multiplication $A^c(:,s)\cdot B^r(s,:)$. The offline version of the double-checksum scheme can only correct a single error in full execution, while the online version, which corrects a single error for \textit{each} step of the outer-product update, can handle multiple errors for the whole program. Later in 2011, Ding and colleagues extended the outer-product GEMM with ABFT to GPUs \cite{ding2011matrix}. To further hide the checksum-related memory latency introduced by ABFT, Zhai et al. designed fused compute kernels for GEMM on AVX-512-enabled CPUs \cite{zhai2021ft}. On GPUs, Kosaian and Rashmi presented an initial trial of GEMM with fault tolerance that is capable of only detecting rather than correcting errors \cite{kosaian2021arithmetic}. Although they managed to fuse the memory latency of checksum calculations with the data movement of the original GEMM operations, explicitly established extra memory spaces must be allocated beforehand to correct a detected error, which leads to inevitable modifications on the dependent source codes. To remedy this deficit, in this paper, we present the first ever fully-fused GEMM with online fault tolerance, which can efficiently both detect and correct computing errors.

\subsection{Kernel Fusion}

Kernel fusion is a performance optimization technique that has been extensively explored in the context of both CPUs and GPUs. The primary goal of kernel fusion is to combine multiple computational kernels into a single, more efficient kernel, thereby reducing the overall overhead associated with kernel launches, memory transfers, and other performance bottlenecks. By exploiting data locality, minimizing memory access, and enhancing cache utilization, kernel fusion has been demonstrated to yield significant performance improvements in a variety of applications and computational domains.

Several works have investigated the potential benefits of kernel fusion on CPUs and GPUs across a range of applications. For example, Chen et al. \cite{chen2018tvm} proposed a compilation approach to automatically fuse numerical kernels in linear algebra and deep learning operations on CPUs. In the context of GPUs, Jia et al. \cite{jia2014caffe} integrated kernel fusion into the popular deep learning framework Caffe, while Rhu et al. \cite{rhu2018compressing} introduced a technique for fusing convolutional layers in deep neural networks to reduce memory footprint and improve performance. Dao et al. \cite{dao2022flashattention} and Zhai et al. \cite{zhai2022bytetransformer} presented fused schemes that significantly boosted the performance of the cornerstone algorithm of the transformer model, multi-head attention, on GPUs. Kernel fusion has also demonstrated benefits in other domains, such as homomorphic encryption \cite{zhai2022accelerating} and scientific computing \cite{chen2023high}. These studies underscore the effectiveness of kernel fusion as a performance optimization strategy for various applications on both CPUs and GPUs, emphasizing its importance in the pursuit of high-performance computing. In this paper, we present new insights into performance optimizations for fault-tolerant GEMM on GPUs by exploiting the kernel fusion strategy.

\section{Optimizing SGEMM without fault tolerance}
\label{sec:design_and_optimizations}

An efficient fault-tolerant GEMM implementation necessitates fusing the memory footprint of ABFT with the original GEMM operations. This suggests that an efficient GEMM codebase is the foundation for instrument-fused memory operations of ABFT. As cuBLAS GEMM is closed-source, implementing a high-performance SGEMM from scratch becomes inevitable. In this section, we detail the step-wise optimizations of single-precision GEMM (SGEMM) from a naive baseline to a performance that surpasses the closed-source cuBLAS by leveraging shared memory, vectorized memory access, and prefetching. We then present our automatic code generation strategy to generalize the performance to a broad range of input matrix shapes.

\begin{figure}[ht]
    \centering
    \includegraphics[scale=0.16]{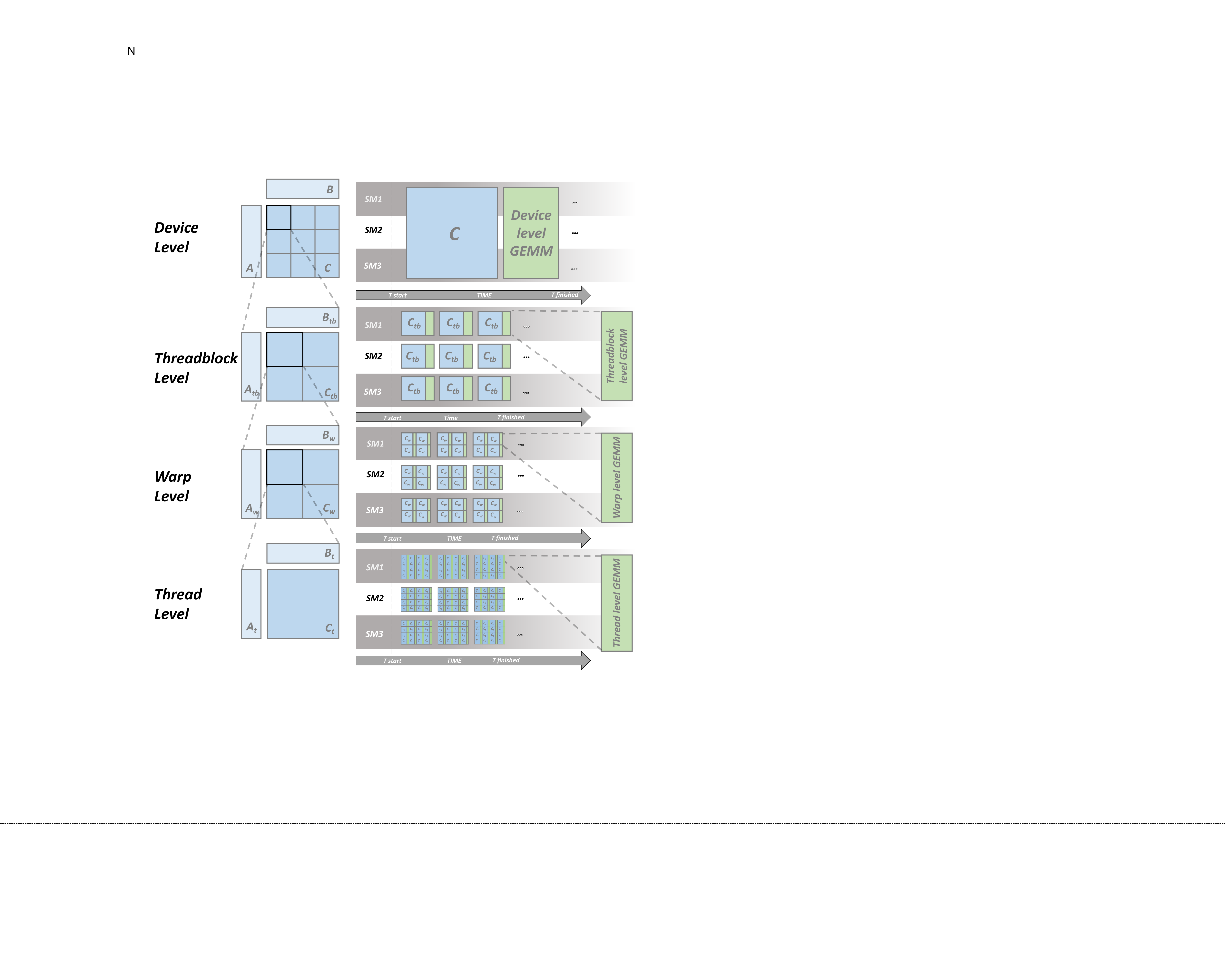}
    \caption{A hierarchical pipeline of the GEMM GPU implementation.}
    \label{fig:sgemm_bird}
\end{figure}

\subsection{Step-wise Optimizations for SGEMM}
\label{subsection:step_wise}
Figure \ref{fig:sgemm_bird} presents a bird's-eye view of SGEMM on GPUs. At the device-level, SGEMM performs $C += A\cdot B$, where $A$, $B$, and $C$ are $M\times K$, $K \times N$, and $M \times N$ matrices, respectively. At the threadblock-level, a threadblock accumulates $C_s += A_s\cdot B_s$, where $A_s$, $B_s$, and $C_s$ are $m_s\times k_s$, $k_s \times n_s$, and $m_s \times n_s$ matrices, respectively. At the warp-level, a warp accumulates $C_w += A_w\cdot B_w$, where $A_w$, $B_w$, $C_w$ are $m_w\times k_w$, $k_w \times n_w$, and $m_w \times n_w$ matrices, respectively. At the thread-level, a thread accumulates $C_t += A_t\cdot B_t$, where $A_t$, $B_t$, and $C_t$ are $m_t\times k_t$, $k_t \times n_t$, and $m_t \times n_t$ matrices, respectively. In this subsection, we present the optimization techniques of GEMM from scratch, detailing the benefits and theoretical underpinnings of each step to unveil the black box of the closed-source cuBLAS library, constituting a foundational codebase for further fault-tolerant GEMM optimizations.

\subsubsection{Naive implementation}


We start presenting the step-wise optimizations from the naive baseline. Here the workload of $C+=AB$ is partitioned in both the row and column dimensions such that each threadblock computes a tile $C_{tb}$ of $C$, namely  $ C_{tb} += A_{tb}\cdot B_{tb}$. Accordingly, each thread calculates an element, $C_{tb}[i,j]$ +=$A_{tb}[i, k]$$B_{tb}[k, j]$, where $k=0,1, \cdots K-1$. The result of $C_{tb}[i,j]$ is accumulated in a register throughout the computation. This baseline variant results in a performance of 611 GFLOPS on average for square matrix inputs ranging from 1024$^2$ to 6144$^2$ on an NVIDIA Tesla T4 GPU.

\subsubsection{Threadblock-level Tiling}

The naive baseline approach suffers from intensive memory access with a complexity of $O(n^3)$ in the slow global memory and does not utilize any data reuse in the faster shared memory. In this improved variant, a tiling strategy is introduced to enable data reuse in the GPU-shared memory, reducing the cost of expensive global memory latency in GEMM operations. The process begins by loading tiles $A_{tb}$ and $B_{tb}$ from the global memory to the shared memory. The threadblock is then synchronized to ensure that all threads finish storing their data. Each thread computes $C_t += A_t \cdot B_t$ by loading elements $A_{tb}[i,k]$ and $B_{tb}[k,j]$ from the shared memory into registers $A_t$ and $B_t$. Our evaluations show that this step results in an average improvement of 11.3\%, with a performance boost to 679 GFLOPS.

\subsubsection{Thread-level Tiling}

Considering the nature of GEMM algorithms, where $O(n^3)$ computations are conducted among $O(n^2)$ matrix elements, we seek further data reuse at the register level in each thread, namely thread-level tiling at this step. Rather than assigning one target element in $C_{tb}$ to each thread, thread-level tiling introduces more workloads to each thread which leads to more aggressive register-level data re-use in registers. The thread-level tiling step better leverages the architectural resources at the register level to significantly reduce memory access in both global and shared memory. We explore and benchmark a series of thread-level tiling variants by assigning $4$, $16$, and $64$ elements of the $C$ matrix to each thread. The SGEMM performance increases by up to $4.62\times$ from the previous step, pushing the GFLOPS to $3822$.

\subsubsection{Warp-level Tiling}

There is another architectural level where the tiling strategy can be employed --- the warp level. The 32 threads in a warp are either partitioned to $4$-by-$8$ or $8$-by-$4$, corresponding to either $4$ or $8$ threads assigned to the row or column dimension. For a $4$-by-$8$ warp-level tiling setup and a $128$-by-$128$ threadblock-level tiling with $256$ threads ($8$ warps) statically initialized in each threadblock, each warp is responsible for a $64\times32$ sub-matrix of $C$ while each thread is responsible for a $16\times4$ sub-block. Since warps are partitioned in a $4$-by-$8$ manner, the first $4$ threads share the first one-eighth column tile of the warp sub-matrix, the next $4$ threads take the second one-eighth column tile of the warp sub-matrix, and so on. A delicate warp-level tile, accordingly, will benefit the memory access at the shared-memory level. In NVIDIA GPUs, when different threads attempt to access the same address in the shared memory, these memory accesses will be reduced to a single one by the hardware. This warp-level tiling strategy, which is also free from a shared memory bank conflict, significantly benefits the memory latency at the shared memory level, increasing the performance to $4331$ GFLOPS --- a $13\%$ speedup from the previous step.

\subsubsection{Vectorized load/store}


Rather than accessing the global and shared memory in a scalar manner, where each memory transaction involves a single data element, we can employ a vectorized memory access to better saturate the memory bandwidth. By adopting 128-bit vectorized memory accesses, we observe a moderate improvement from the previous $4331$ GFLOPS to the current $4381$ GFLOPS on average.


\subsubsection{Prefetching: shared memory to registers}


To further hide the memory latency of accessing the shared memory, we double the register utilization and pipeline the shared memory accesses with the GEMM computations, namely prefetching at this step. The doubled register utilization breaks the original load-compute dependency, enabling the data to be computed to be loaded into registers in the previous iteration. During each iteration, a thread prefetches two fragments, $A_{tb}[i:i+8, k+1]$ and $B_{tb}[k+1, j:j+8]$. The thread then performs the accumulation $C_t[0:8, 0:8] += A_t[0:8] \times B_t[0:8]$, using the fragments $A_{tb}[0:8, k]$ and $B_{tb}[k, 0:8]$ currently in the registers. After the accumulation, the prefetched fragments are loaded into the registers, $A_t[0:8]$ and $B_t[0:8]$. This optimization results in a 5.6\% improvement in performance, with GFLOPS increasing from 4381 to 4625.

\subsubsection{Prefetching: global memory to shared memory}


The technique of prefetching can also be used for the transfer of data between global memory and shared memory by doubling the size of the shared memory buffers. This ensures that all the data needed in shared memory is stored in the correct position in the previous iteration. For instance, the prefetched global memory tiles $A[i:i+128, k+8:k+2\times 8], B[k+8:k+2\cdot 8, j:j+128]$ are loaded into $A_{tb}[0:128, 0:8]$ and $B_s[0:8, 0:128]$ after the current accumulation $C_{tb} += A_{tb}B_{tb}$ is complete. The accumulation uses the tiles $A[i:i+128, k:k+8]$ and $B[k:k +8, j:j+128]$ that are currently in the shared memory. The $\mathtt{\_\_syncthreads()}$ function is used to ensure that all threads have finished the accumulation. By doubling the shared memory buffers from $A_{tb}[0:128, 0:8]$ and $B_{tb}[0:8, 0:128]$ to $A_{tb}[0:128, 0:2\times 8]$ and $B_{tb}[0:2\times8, 0:128]$, the prefetched tiles can be stored directly in the currently unoccupied portion of the shared memory without waiting for the accumulation to finish and the $\mathtt{\_\_syncthreads()}$ function. At this step, the SGEMM performance is further improved by $0.6\%$, increasing to $4654$ GFLOPS.


\begin{figure}[ht]
    \centering
    \includegraphics[scale=0.21]{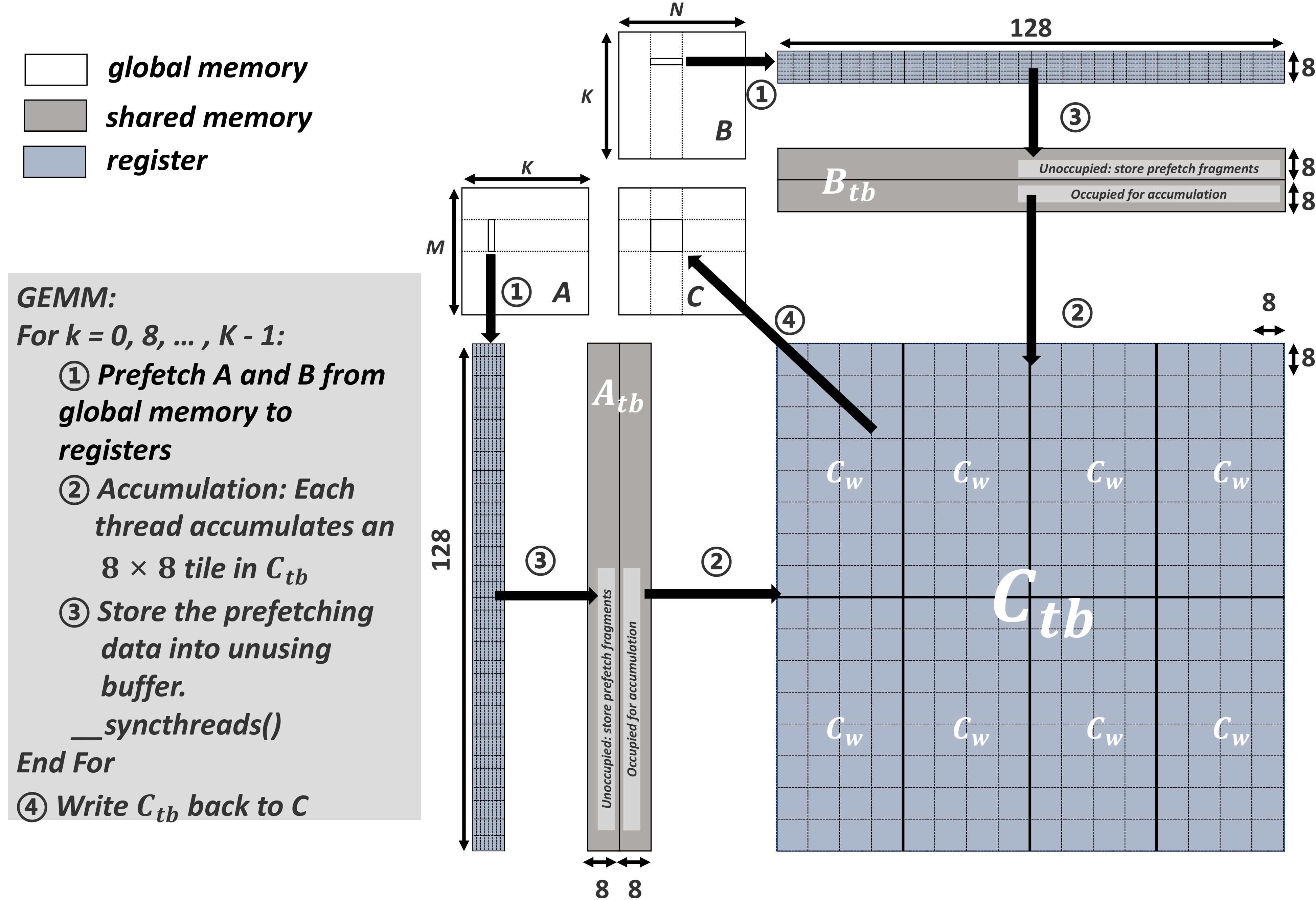}
      \caption{Overview of the optimized SGEMM kernel. }
    \label{fig:sgemm}
    \vspace{-4mm}
\end{figure}

\subsection{Automatic Code Generation for SGEMM}
\label{sec:codegen_sgemm}


The static hard-coded parameter selection mechanism can degrade performance when applied to broader input shapes. For instance, when using a 128-by-128 2-D partitioning method for the output matrix, the performance is satisfactory for larger matrices. However, the same parameters result in insufficient active threadblocks being launched during runtime for smaller input matrices, which under-utilizes the GPU. To address this, we propose a more flexible parameter selection mechanism in our work. However, integrating a series of hard-coded kernels incurs significant development costs. As such, we utilize a templatized approach for our SGEMM kernels, which involves feeding semi-empirically selected kernel parameters according to the input shapes that generate high-performance parameterized kernels at runtime. This online code generation scheme allows us to generalize the superior performance of SGEMM on square matrices to irregularly shaped inputs while keeping development costs low.

\begin{table}[ht] \centering
\caption{SGEMM kernel parameter setup on a Tesla T4 GPU.}
\begin{tabular}{l
    S[table-format=3] 
    S[table-format=3] 
    S[table-format=3] 
    S[table-format=3] 
    S[table-format=3] 
    S[table-format=3] 
    S[table-format=3]
    }
\toprule
                   & {$m_{tb}$}       & {$n_{tb}$}       & {$k_{tb}$}  & {$m_w$} & {$n_w$} & {$m_t$} & {$n_t$} \\ 
\midrule
small              & {$16$}          & {$16$}              & {$16$}    & {$8$} & {$16$}    & {$2$}   &  {$2$} \\ 
medium              & {$32$}          & {$32$}             & {$8$}    & {$16$}  & {$32$}    & {$4$}  & {$4$} \\ 
large              & {$64$}         & {$64$}             & {$8$}      & {$32$} & {$64$}    & {$8$} & {$8$}\\ 
tall and skinny    & {$32$}         & {$128$}             & {$8$}      & {$16$}  & {$64$}   & {$4$} &{$8$} \\  
huge          & {$128$}         & {$128$}             & {$8$}      & {$32$} & {$64$}    & {$8$} &  {$8$} \\ 
\bottomrule
\end{tabular}
\label{tab:kernel_size}
\end{table}

\subsubsection{Code Generation Strategy}

Besides following the step-wise SGEMM optimization, the code generation scheme takes 7 parameters as input and generates a corresponding high-performance SGEMM kernel. The kernel parameters are $n_{tb}, m_{tb}, k_{tb}, m_w, n_w, m_t$, and $n_t$. These parameters correspond to the tile sizes in threadblock-level ($tb$), warp-level ($w$), and thread-level ($t$), respectively. In the code generation template, the memory operations are well-designed to avoid bank conflicts. Other parameters are directly determined by the tile sizes, e.g. double buffer size,  and data type of vectorized load/store. Figure \ref{alg:codegen_pseudo} shows the code generation template of SGEMM.

\subsubsection{Kernel Parameters}
We generated SGEMM kernels with various parameters using the code generation template. Through empirical analysis, we identified the four best kernels for input shapes of $1\sim128$, $128\sim256$, $256\sim512$, and $>512$, respectively. These kernels are named small, medium, large, and huge. Additionally, we developed a kernel optimized for irregular input shapes, named tall and skinny. Here we adopt a parameter selection scheme inspired by a series of works based on NVIDIA CUTLASS \cite{nv-cutlass, huang2020strassen}. Table \ref{tab:kernel_size} summarizes the details of the corresponding kernel parameters.

\begin{figure}[ht]
    \centering
    \includegraphics[scale=0.19 ]{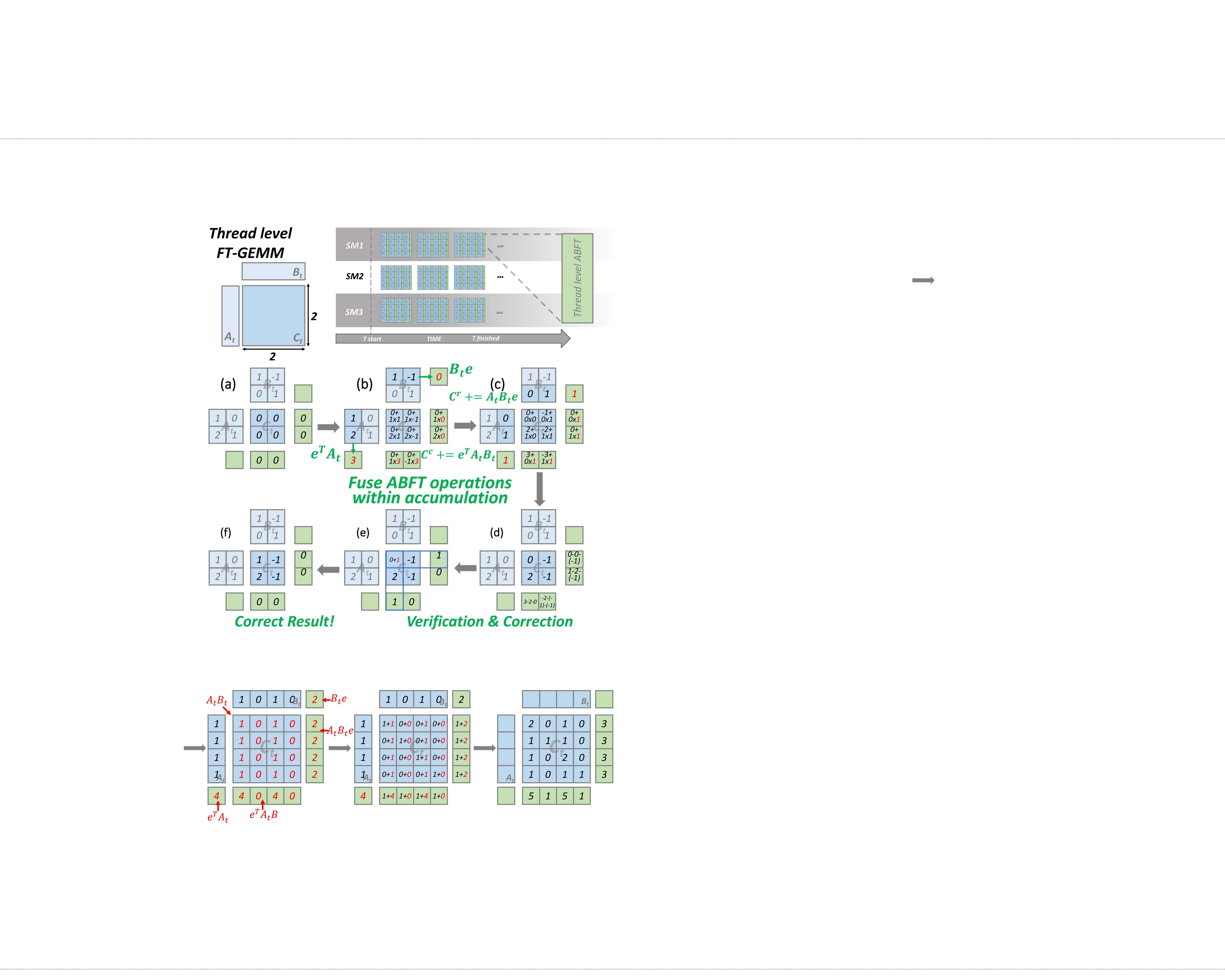}
    \caption{Thread level FT-SGEMM.}
    \label{fig:toy_thread_abft}
\end{figure}
\section{Optimizing SGEMM With Fault Tolerance}
\label{sec:fault_tolerant}

The self-implemented, highly-efficient SGEMM kernel provides us with a high-performance codebase to build lightweight fault-tolerant schemes within compute kernels. In this section, we explore fused ABFT schemes for GEMM with respect to three different hardware levels --- the thread level, warp level, and thread block level. 

\subsection{Fault Model}

We elaborate on our fault model before delving into the fault tolerance design. Our study is centered on the detection and correction of errors at computing units that can affect the results of the final output matrix. We do not direct our attention to faults in the memory, as these are relatively more resilient under ECC. To address the compute errors at run-time,  we design our fault-tolerant scheme under a single-event upset (SEU) assumption, i.e., there is only one soft error in each error detection and correction period. The SEU assumption, which is a widely used fault model in many research works \cite{reis2005swift, zhai2021ft, ding2011matrix, wu2014ft}, is valid because of the low occurrence rate of multiple soft errors caused by short fault detection intervals.


\subsection{Exploring Various ABFT SGEMM schemes}

\subsubsection{Thread-level ABFT SGEMM}



In our initial trial, we aim to implement the fault-tolerant functionality at the thread level. At the thread level, ABFT-related operations can be fused within the thread-level accumulation without additional intra-warp or intra-threadblock communications.

Figure \ref{fig:toy_thread_abft} demonstrates the process of thread-level FT-GEMM detecting and correcting a fault. In Figure \ref{fig:toy_thread_abft}(a), a thread initializes $4$ registers for $C_t$ (marked in blue) and $6$ registers for checksum (marked in green). $A_t$ and $B_t$ are stored in shared memory (marked in white). In Figure \ref{fig:toy_thread_abft}(b) and (c), the thread loads fragments of $A_t$ and $B_t$ to perform the matrix multiplication. The checksums of the fragments and $C_t$ are updated simultaneously, namely $e^TA_t$, $B_te$, $C^c+=e^TA_tB_t$, and $C^r+=A_tB_te$.  In Figure \ref{fig:toy_thread_abft}(d), $C_t$ is verified using the checksum. In Figure \ref{fig:toy_thread_abft}(e), a fault is detected and corrected. The fault location is determined by relative positions in two checksums. On the other hand, the correction value is determined by the offset in the checksums. Finally, we get the fault-free $C_t$ in Figure \ref{fig:toy_thread_abft}(f).

The thread-level ABFT scheme, though it does not come with the cost of intra-warp or intra-threadblock communications, leads to a considerable computational overhead. This is because the standard GEMM has a cubic computational complexity, whereas that of the checksum encoding is quadratic. The larger the protected matrix is, the lower the overhead from the checksum encoding will be. However, due to limited register resources, the thread-level ABFT handles only small matrices (e.g., $8\times8$), leading to significant overhead. In practice, on an NVIDIA Tesla T4 GPU, thread-level ABFT results in a 25\% average overhead for square matrices ranging from $1024^2$ to $6144^2$.

\subsubsection{Warp-level ABFT SGEMM}
\begin{figure}[ht]
    \centering
\includegraphics[scale=0.17]{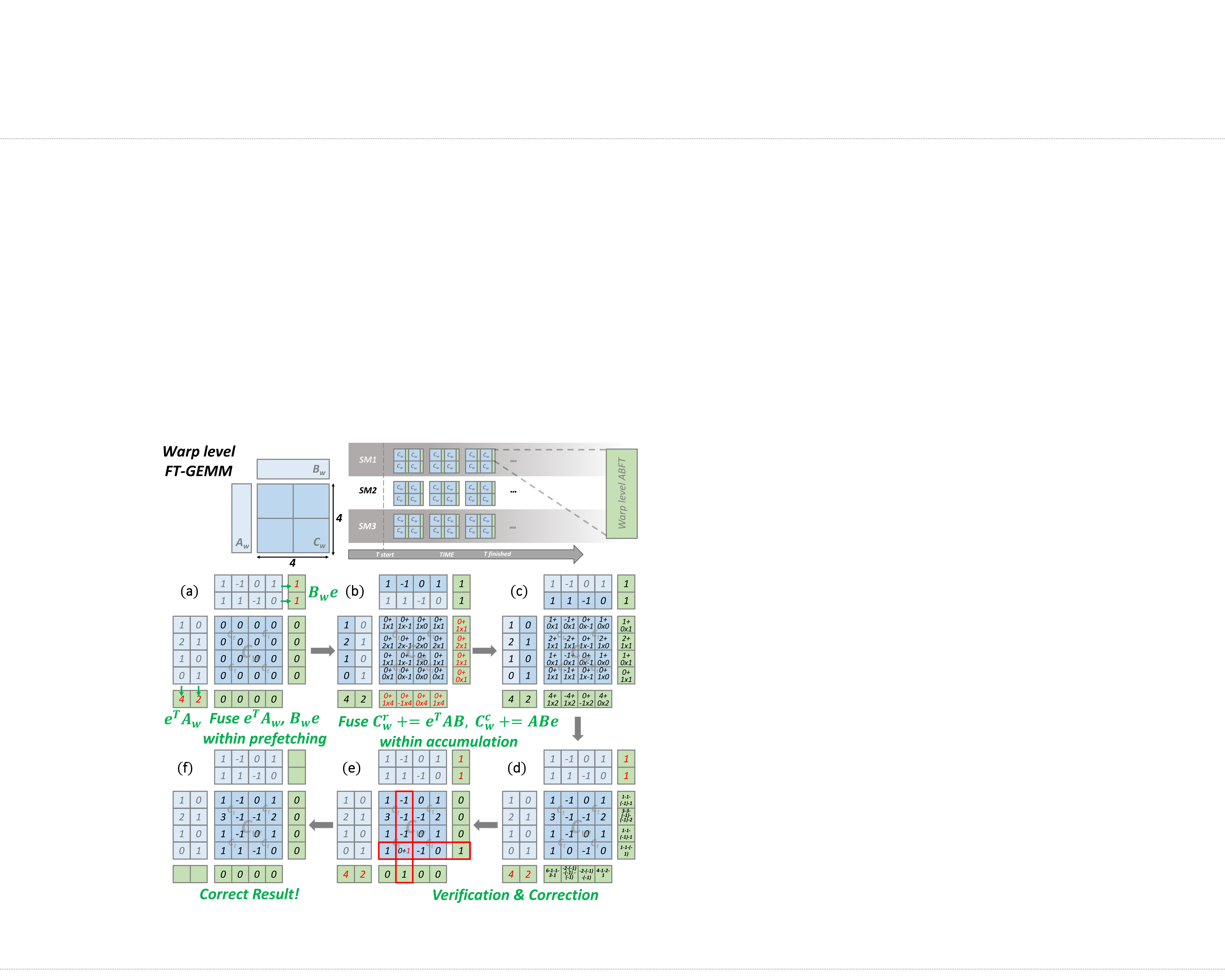} 
    \caption{Warp Level ABFT. A warp contains $4$ threads for simplicity.}
    \label{fig:toy_warp_abft} 
\end{figure}




Although the thread-level ABFT does not increase memory operations, the encodings of ABFT bring a $(4n_t) / (2n_t^2) = 2/n_t$ increase in total computations, where $n_t$ is the tile size of a thread. This results in a $25\%\sim100\%$ increase in computations according to Table \ref{tab:kernel_size}. Considering the computational intensity of GEMM, the increase in computations results in a non-negligible overhead. Hence, it is a good choice to increase the ABFT encoding unit from thread to warp. Ideally, warp-level ABFT will reduce the increment in computations to $5\%$ while introducing several warp-level communications. The communication includes a warp-level reduction for the encodings of $e^TA_w$ and $B_we$, distribution of $e^TA_w$ and $B_we$, and gathering of $C_w^c$ and $C_w^r$ from different threads. First, we find the warp-level reduction can be fused with the prefetching of $A_{tb}$ and $B_{tb}$  (step 3 in Figure \ref{fig:sgemm}). Each warp holds one or more fragments of $A_w$ and $B_w$. Hence, $e^TA_w$ and $B_we$ can be computed with $2\sim5$ calls of warp-level reduction primitives. The primitive exchanges data at the register level without additional latency for synchronization. Next, we reuse the idle part of the double buffer to distribute $e^TA_{w}$ and $B_{w}e$ without the requirements for extra memory space. After that, $C_w^c$ and $C_w^r$ are updated simultaneously with GEMM. During verification, $C_w^c$ and $C_w^r$ are first gathered from threads into the shared memory and then compared with $C_w$. Finally, each thread checks $C_w^c$ and $C_w^r$ to verify and correct the result.

\begin{figure}[ht]
    \centering
\includegraphics[scale=0.13]{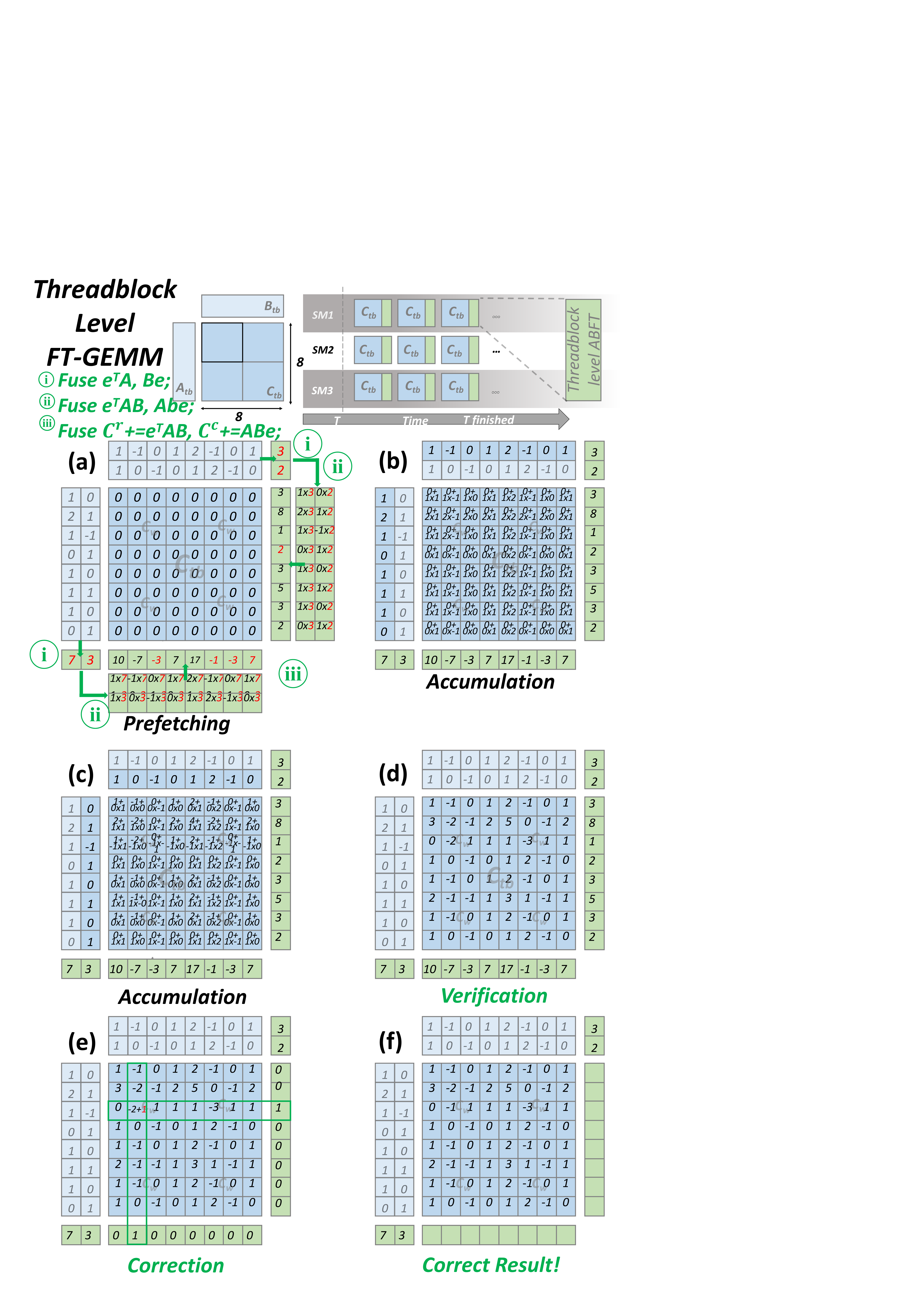} 
    \caption{Threadblock Level ABFT.}
    \label{fig:toy_tb_abft} 
\end{figure}

Figure \ref{fig:toy_warp_abft} demonstrates a toy example of warp-level ABFT. In Figure \ref{fig:toy_warp_abft}(a), the encodings of $e^TA_w$ and $B_we$ are computed and stored in shared memory (marked in white). In Figure \ref{fig:toy_warp_abft}(b) and (c), each thread loads fragments of $A_w$ and $B_w$ to perform the matrix multiplication. The encodings $e^TA_w$ and $B_we$ are loaded simultaneously for the following updates: $C_w^c+=e^TA_wB_w$ and $C_w^r+=A_wB_we$.  In Figure \ref{fig:toy_warp_abft}(d), $C_w^r$ and $C_w^c$ are first loaded into the shared memory and then compared with $C_w$. In Figure \ref{fig:toy_warp_abft}(e), a fault is detected and corrected. Finally, we get the fault-free $C_w$ in Figure \ref{fig:toy_thread_abft}(f). Although warp-level ABFT has a low computation overhead, it requires two additional shared memory reads whenever $C_w$ is updated. Although the read operations do not require synchronization, they still limit the efficiency of the GEMM. Hence, we propose threadblock-level ABFT, which further decreases the additional computation and fuses all communications within the prefetching to minimize the overhead.


\subsubsection{Threadblock-level ABFT SGEMM}


To further hide the overhead of ABFT,  we apply ABFT encodings to the threadblock-level. To avoid additional latency during GEMM accumulation, we fuse all encodings with the prefetching stage. Figure \ref{fig:toy_tb_abft} demonstrates a toy example of the threadblock-level ABFT scheme. As shown in Figure \ref{fig:toy_tb_abft}(a), all ABFT encodings are fused with the prefetching stage. With a well-designed prefetching strategy, each element in $e^TA$ and $Be$ can be obtained with a warp-level reduction without extra global read operations. After that, our prefetching strategy enables the encoding of $e^TAB$ and $ABe$ to be available within a thread instead of a threadblock-level communication. Finally, the target checksums $C^r$ and $C_c$ are updated through a threadblock-level reduction. In Figure \ref{fig:toy_tb_abft}(b-c), the GEMM accumulation is performed without additional operations.  Figure \ref{fig:toy_tb_abft}(d-f) follow the same verification and correction logistics as thread-level ABFT and warp-level ABFT.

\begin{figure}[ht]
    \centering
\includegraphics[scale=0.20]{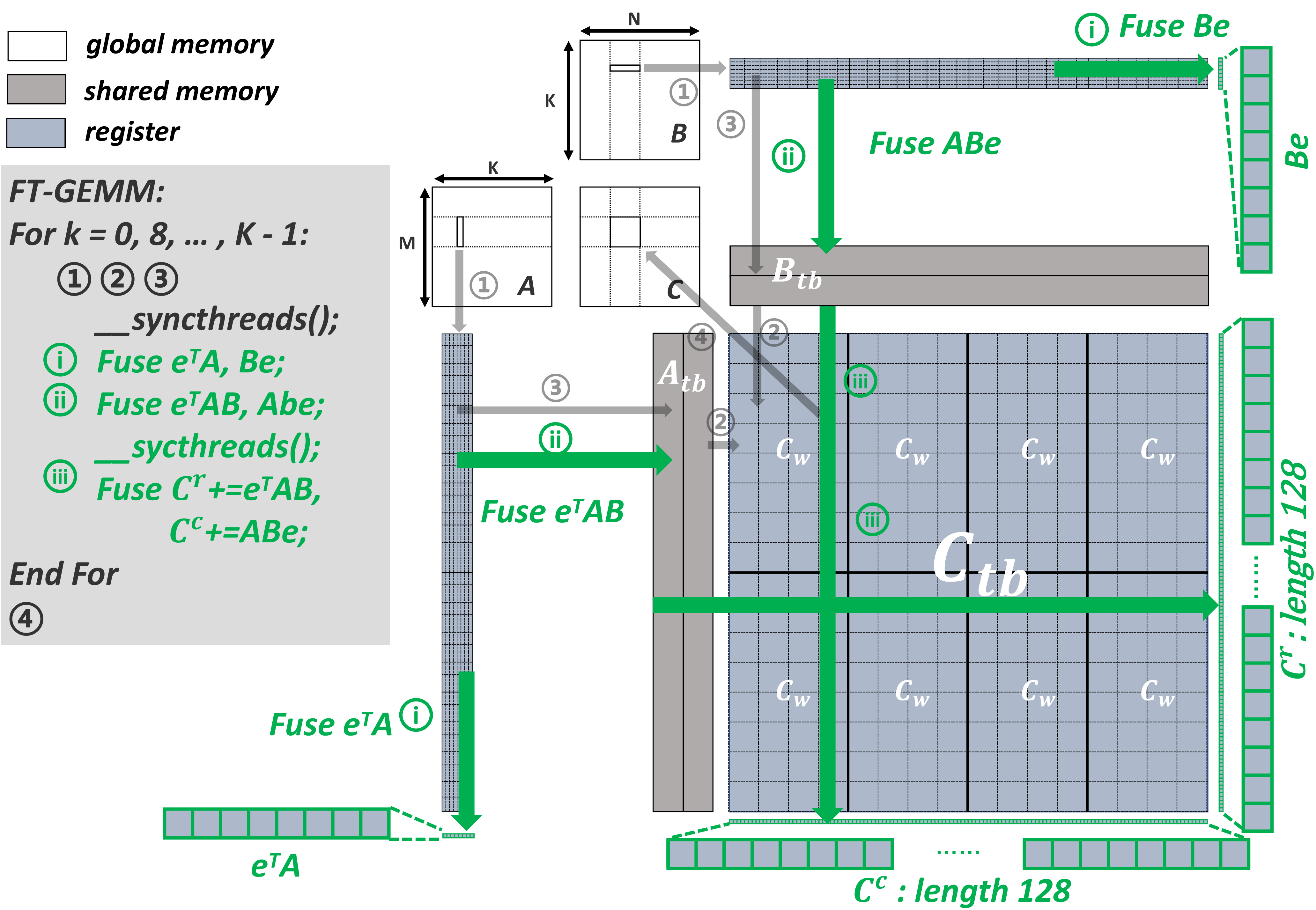} 
    \caption{Threadblock Level ABFT. The parameters of huge kernel in Table \ref{tab:kernel_size} are selected.}
    \label{fig:tb_abft} 
\end{figure}

\begin{figure}[ht]
    \centering
    \includegraphics[scale=0.21]{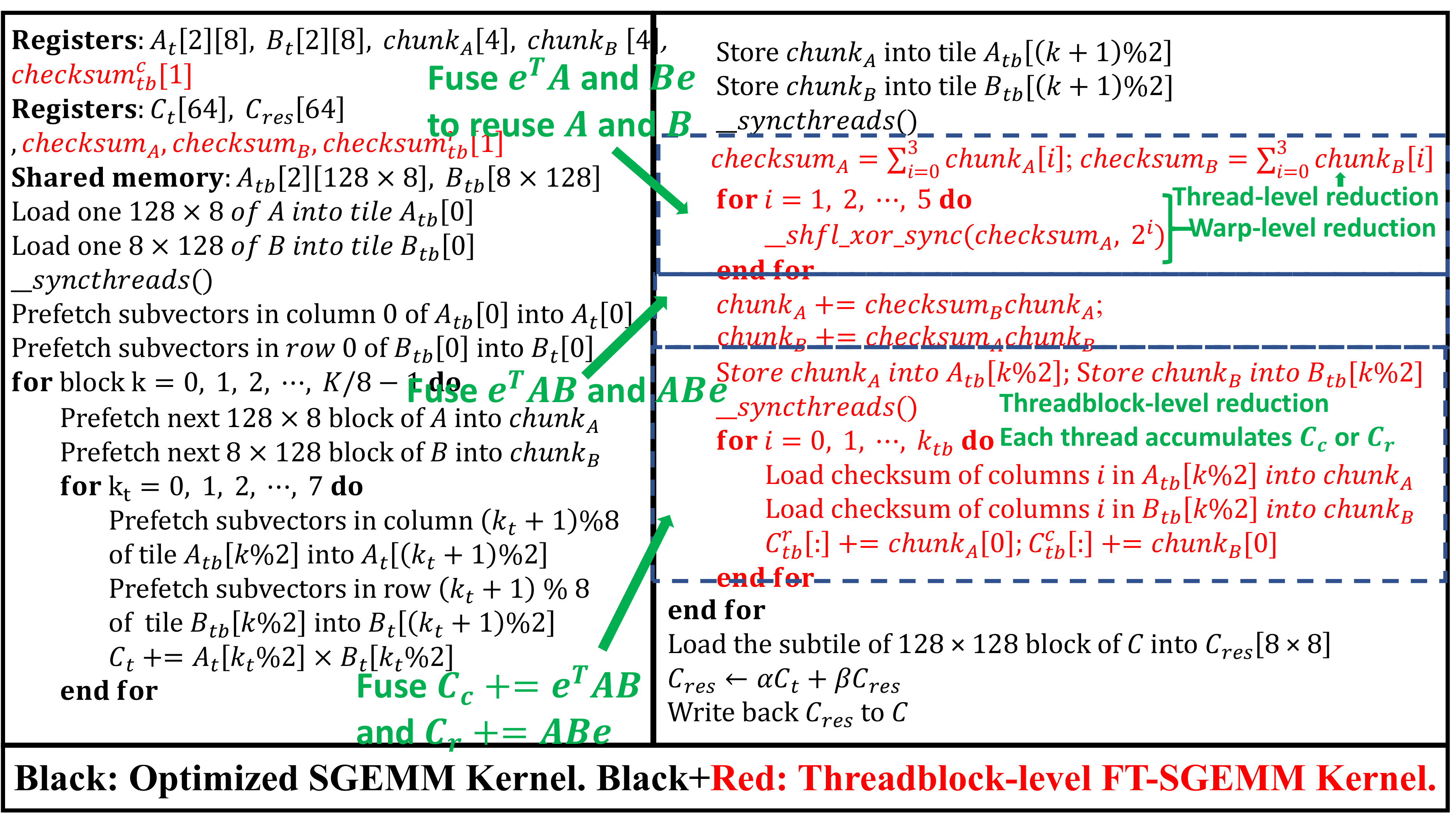}
      \caption{Pseudocode: Optimized SGEMM Kernel vs our hard-coded FT-SGEMM Kernel. The ABFT-related operations are marked in red.}
    \label{alg:nocodegen_pseudo}
\end{figure}

Figure \ref{fig:tb_abft} gives further details on the workflow of threadblock-level FT-SGEMM. The aforementioned checksum encoding and decoding scheme are applied in every threadblock-level outer-product $k$-loop to update the checksum online. Compared to the prior two schemes, threadblock-level ABFT fully leverages the prefetching stage of fragments $A_{tb}$ and $B_{tb}$. Because of the negligible additional computations and the well-designed fusion strategy to hide encoding overhead, the threadblock-level ABFT has the top performance among the other two fused ABFT. Next, we present an automatic FT-SGEMM code generation based on the threadblock-level ABFT. We summarize the pseudo-code of the threadblock-level ABFT and compare it with the ordinary GEMM baseline in Figure \ref{alg:nocodegen_pseudo}. The memory footprints of checksum encoding and online updating, enabled by our highly efficient GEMM kernel built from scratch, are fully fused with the GEMM computations such that the fault-tolerant overhead is minimized.

\subsection{Automatic code generation}



\begin{figure}[ht]
    \centering
\includegraphics[scale=0.21]{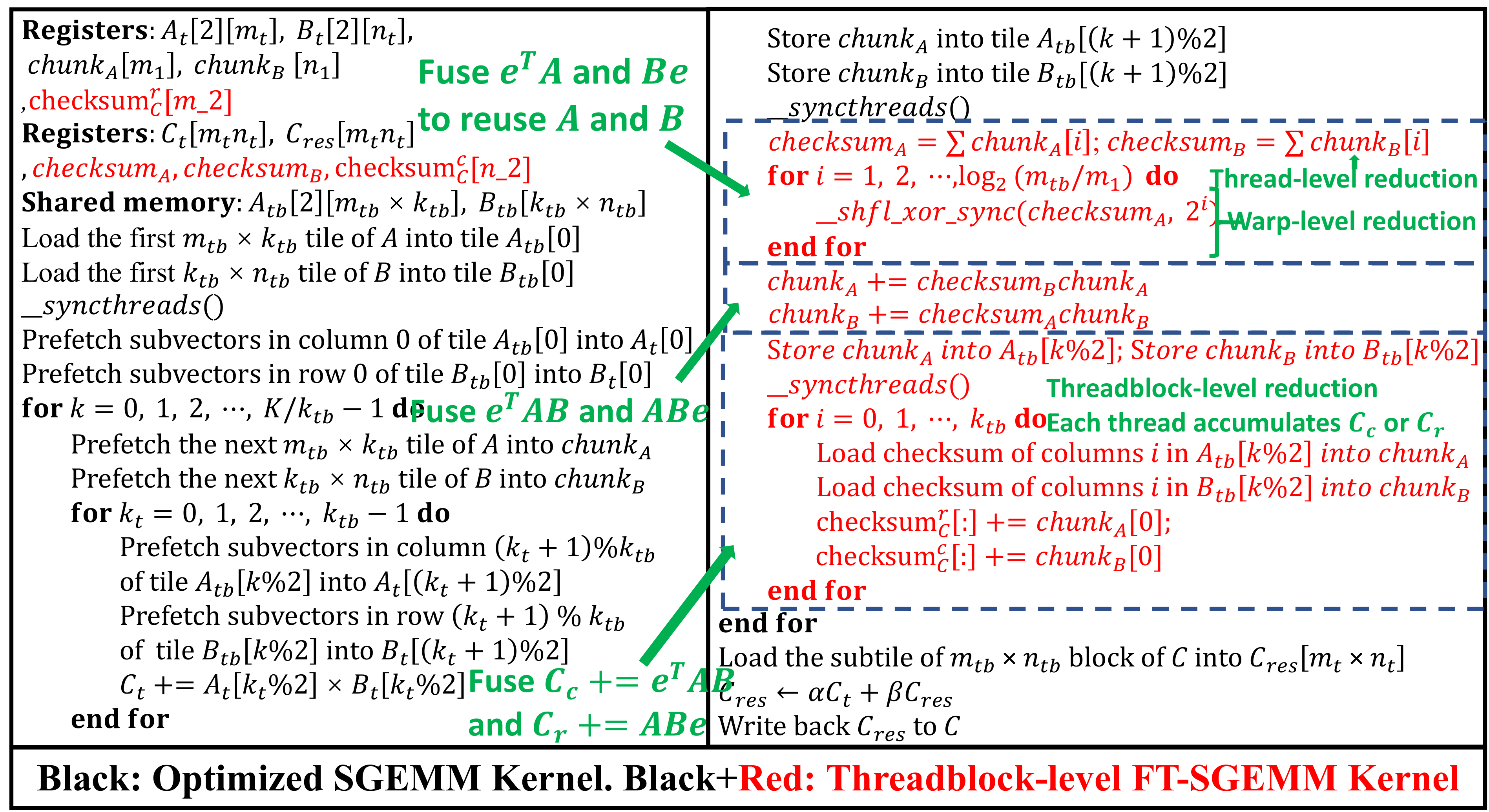} 
    \caption{Pseudocode with code generation: SGEMM vs. threadblock-level FT-SGEMM. The ABFT-related operations are marked in red.}
    \label{alg:codegen_pseudo}
\end{figure}

As discussed in Section \ref{sec:codegen_sgemm}, developing a code generation scheme enables our high-performance kernel to a wide range of input shapes while maintaining a reasonable development cost. Here, we present a code generation strategy for threadblock-level FT-SGEMM. In spite of using the same code generation logistic in Section \ref{sec:codegen_sgemm}, the code generation strategy of FT-SGEMM requires additional designs for ABFT encodings. Considering the wide range of kernel parameters, the ratio between the number of bits in encoding and the number of threads varies. This results in different workloads per thread for a variety of kernel sizes. To maintain superior performance, we first ensure the thread coalesces and avoids bank conflicts. Secondly, we assign workloads equally into different threads. Figure \ref{alg:codegen_pseudo} illustrates a pseudocode for the FT-SGEMM code generation template. The ABFT operations are marked in red. Kernel parameters of the SGEMM keep the settings in Table \ref{tab:kernel_size}.

\section{Performance Evaluation}
\label{sec:results}

We evaluate our optimizations on two NVIDIA GPUs, a Tesla T4 and an A100. The Tesla T4 GPU is connected to a node with two 16-core Intel Xeon Silver 4216 CPUs, whose boost frequency is up to 3.2 GHz. The associated CPU main memory system has a capacity of 512 GB at 2400 MHz. We compile programs using CUDA $\mathtt{11.4}$ with the $\mathtt{-O3}$ optimization flag. The A100 GPU is connected to a node with one 64-core AMD EPYC 7742 CPU with a boost frequency of 3.4 GHz. We evaluate the performance of step-wise SGEMM optimizations, different level ABFT SGEMM strategies, and the generated kernel. We compare our FT-SGEMM with the prior state-of-the-art fault-tolerant GEMM implementation first presented by Ding et al. in 2011 \cite{ding2011matrix}. We also compare our FT-SGEMM against the built-in cuBLAS SGEMM with CUDA $\mathtt{11.6}$. The reported performance data are averaged over tens of runs to minimize fluctuations.

\begin{figure}[ht]
    \centering
\includegraphics[scale=0.24]{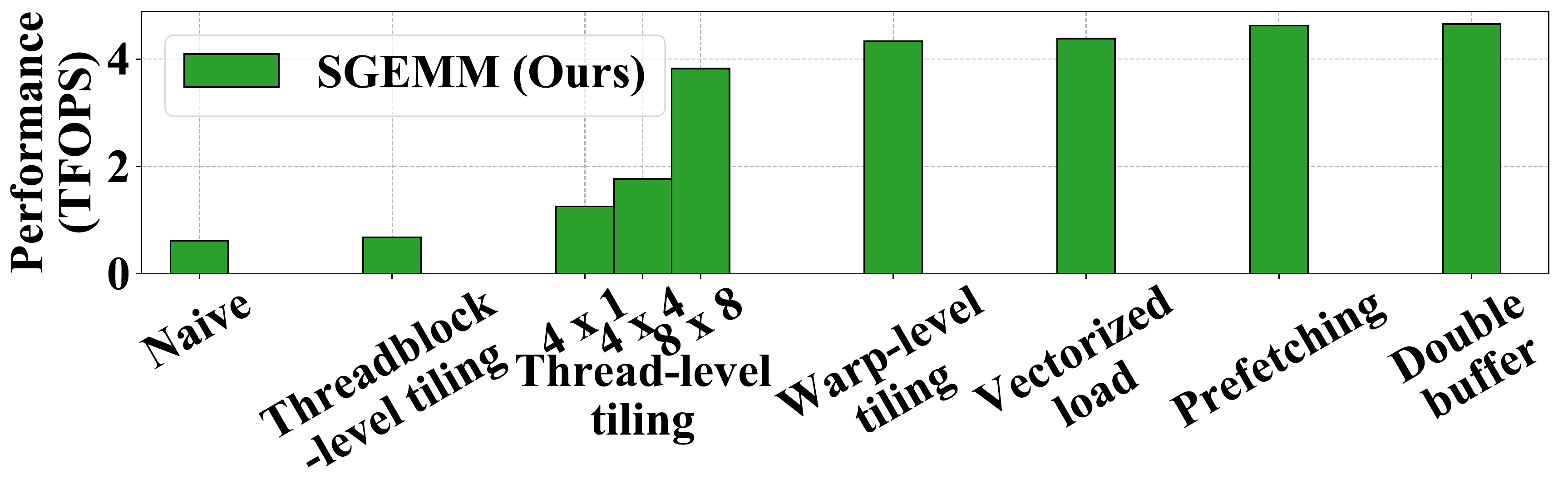} 
    \caption{Step-wise SGEMM optimization.}
    \label{fig:stepwise_sgemm}
\end{figure}
\subsection{Benchmarking SGEMM Without Fault Tolerance}

\subsubsection{Step-wise optimizations for SGEMM}

Figure \ref{fig:stepwise_sgemm} presents the stepwise optimizations of SGEMM without fault tolerance on a NVIDIA Tesla T4 GPU. The baseline performance is $611$ GFLOPS, and with threadblock-level tiling, the performance improves by $11.3\%$ due to the increased data reuse in shared memory. The performance further improves to $3822$ GFLOPS with register-level data re-use. It is worth mentioning that these simple optimizations already outperform cuBLAS SGEMM on the Tesla T4 GPU. Moreover, our proposed pipeline strategy, which overlaps computation with data transactions from global to shared memory and from shared memory to register, achieves $4654$ GFLOPS, which is $7.62$ times faster than the initial variant.

\subsubsection{Automatic code generation for non-FT SGEMM}
\begin{figure}[ht]
    \centering
\includegraphics[scale=0.20]{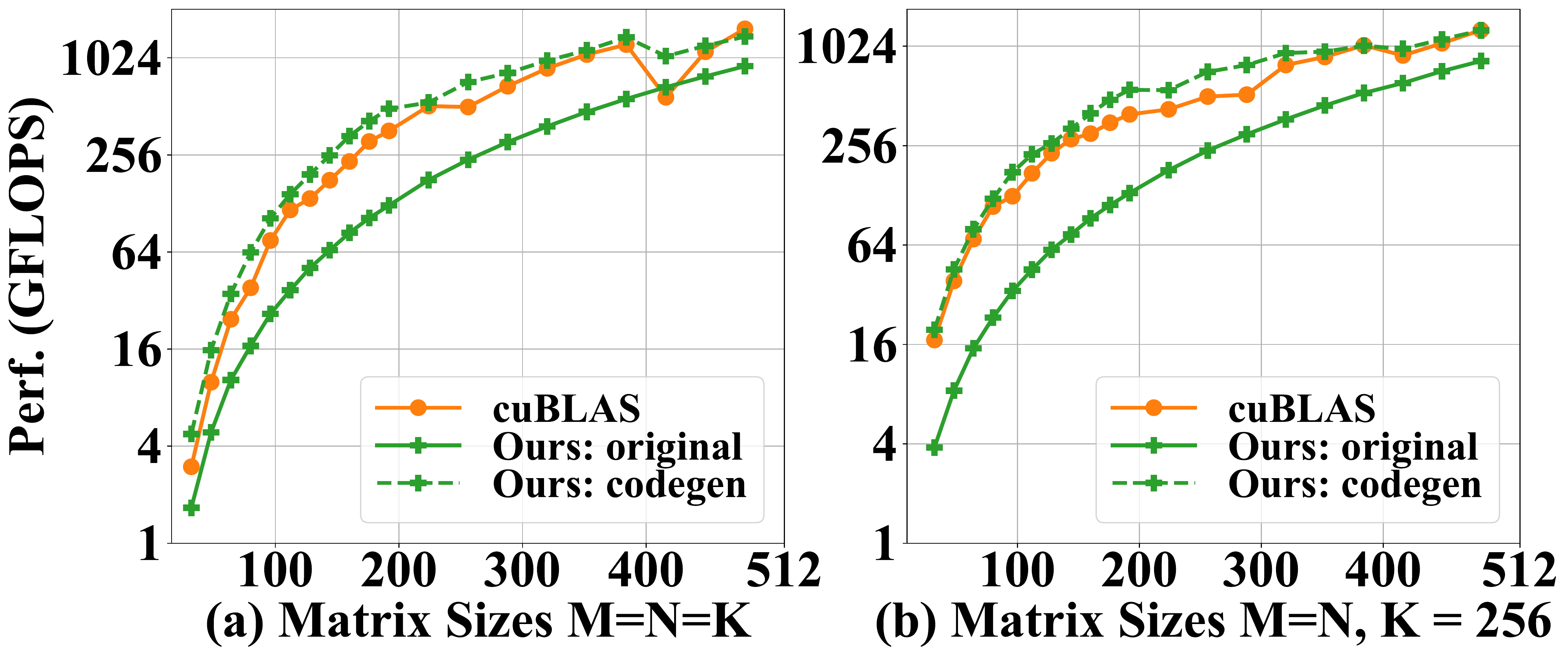}
    \caption{Auto-generated SGEMM kernel performs better for irregular inputs.}
    \label{fig:T4_why_code_gen_noABFT}
\end{figure}

In addition to providing a hard-coded SGEMM implementation, we templatize the SGEMM kernel to generalize its performance from square matrices to a broad range of irregularly shaped inputs without incurring significant human efforts on kernel development. Figure \ref{fig:T4_why_code_gen_noABFT} compares the performance of generated non-FT SGEMM kernels with cuBLAS and original kernels on the same Tesla T4 GPU. By automatically selecting a set of more flexible partitioning parameters at run-time, the generated kernels give an improvement over the hard-coded baseline by up to $230.96\%$. Compared with the closed-source cuBLAS SGEMM, our auto-generated kernels also show a $18.21\%$ speedup on average.

\begin{figure}[ht]
    \centering
\includegraphics[scale=0.19]{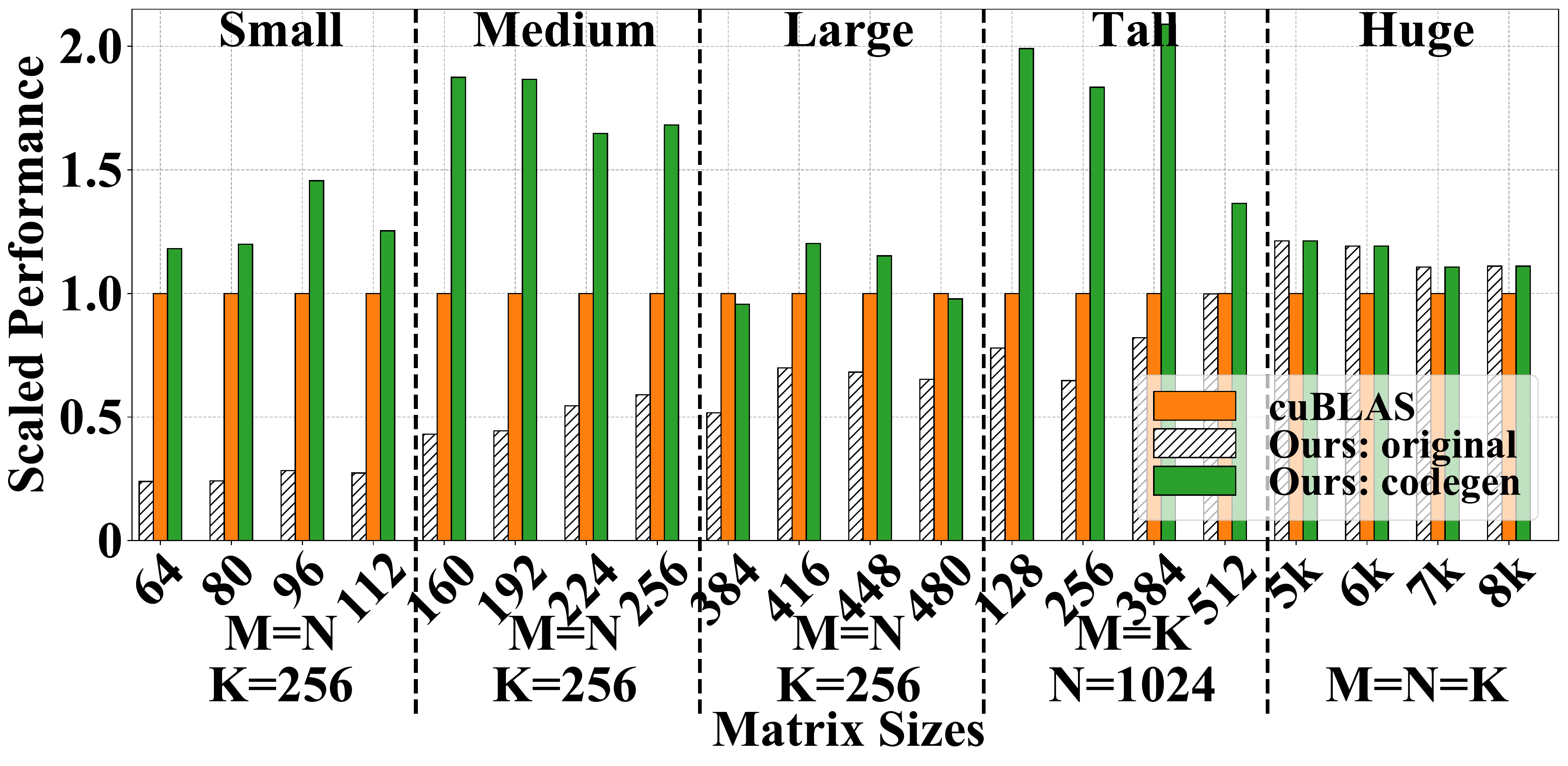} 
    \caption{Performance of generated SGEMM kernels.}
    \label{fig:T4_sgemm_bar_huge}
\end{figure}

We further generalize our template-based code generation strategy from small square matrices to a series of irregularly shaped inputs, namely $\mathtt{small, medium, large, tall}$, and $\mathtt{huge}$ in Figure \ref{fig:T4_sgemm_bar_huge}. By adopting the parameter selection strategy in Table \ref{tab:kernel_size}, our generated SGEMM kernels show a leading performance for various input shapes beyond large square matrices. By fixing the $K$ dimension to $256$ and varying $M$ and $N$ from $64$ to $490$ with a step size of $32$, the code generator takes different kernel parameters and generates the corresponding SGEMM kernel. For small matrices ($M$ and $N$ from $64$ to $112$), our code generator outperforms cuBLAS by an average of 27.23\%. For medium-sized matrices ($M$ and $N$ equal to $160$), the code generator switches to another parameter setup and outperforms cuBLAS by 76.72\%. For larger matrices ($M$ and $N$ $\ge 384$), the code generator again takes different parameters and exceeds cuBLAS by 7.22\%. For wider matrices with $K$ fixed at $1024$, the code generator outperforms cuBLAS by 81.95\%. It is worth mentioning that the parameterized kernels perform significantly better than the hard-coded SGEMM for irregularly shaped matrices, with an average improvement of 160\%.

\subsection{Benchmarking SGEMM With Fault Tolerance}

Our optimizations on SGEMM from scratch, including tiling, register- and shared-memory-level data re-use, pipelining, and template-based flexible kernel parameter selection, enable us to further explore light-weight fault-tolerant schemes for SGEMM.

\begin{figure}[ht]
    \centering
\includegraphics[scale=0.22]{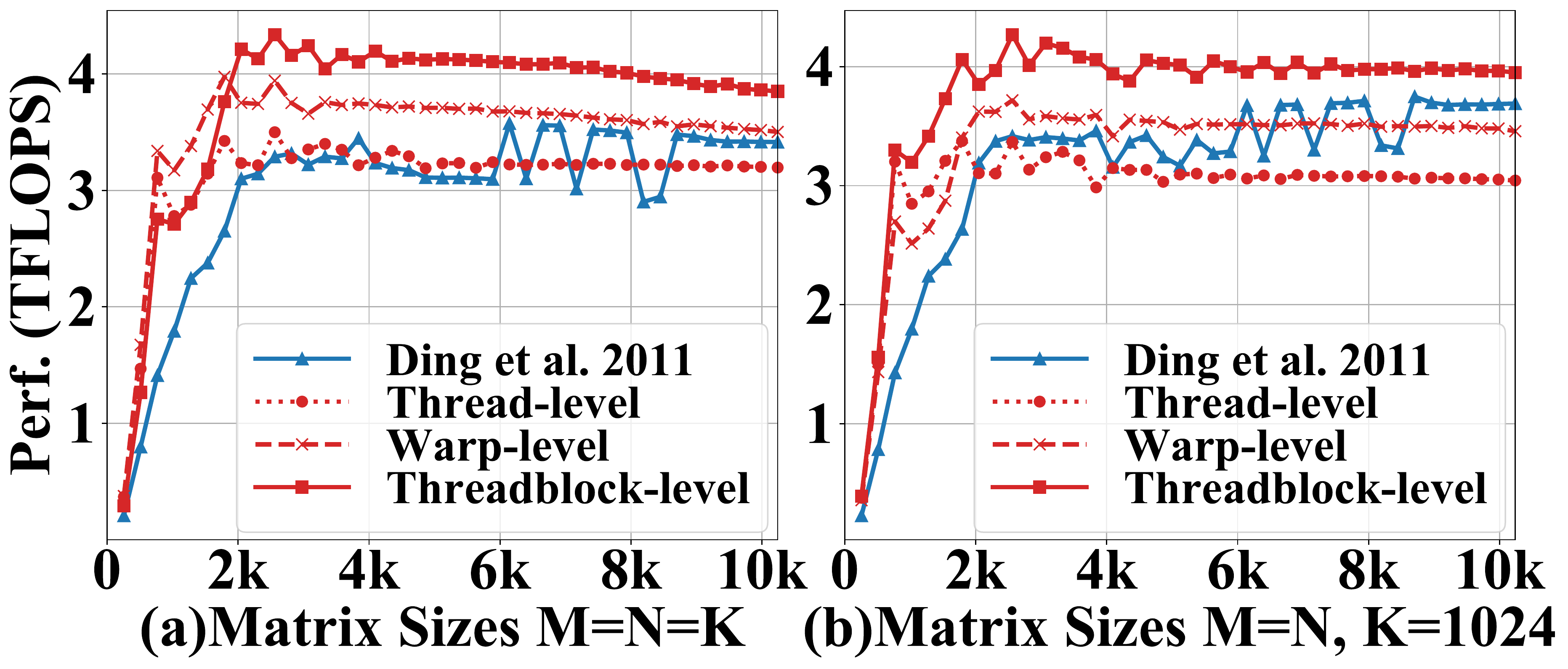} 
    \caption{Different schemes of FT-SGEMM on a Tesla T4 GPU.}
    \label{fig:T4_diff_level_ftsgemm}
\end{figure}

\subsubsection{Benchmarking various schemes for SGEMM with fault tolerance.} 

Figure \ref{fig:T4_diff_level_ftsgemm} benchmarks the performance of our proposed fault-tolerant schemes at the thread, warp, and threadblock levels and compares them against the prior state-of-the-art non-fused baseline. All of our proposed fused ABFT schemes outperform the non-fused FT-SGEMM. Among the 3 different fused kernels, the threadblock-level FT-SGEMM shows the best performance. For the $M=N=K$ case, the threadblock-level FT-SGEMM outperforms the non-fused kernel, thread-level FT-SGEMM, and warp-level FT-SGEMM by $25.98\%$,  $19.55\%$, and $6.49\%$, respectively. For the $M=N, K=1024$ case, the threadblock-level FT-SGEMM outperforms the non-fused kernel, thread-level FT-SGEMM, and warp-level FT-SGEMM by $23.14\%$, $24.85\%$, and $13.88\%$, respectively.

\begin{figure}[ht]
    \centering
\includegraphics[scale=0.195]{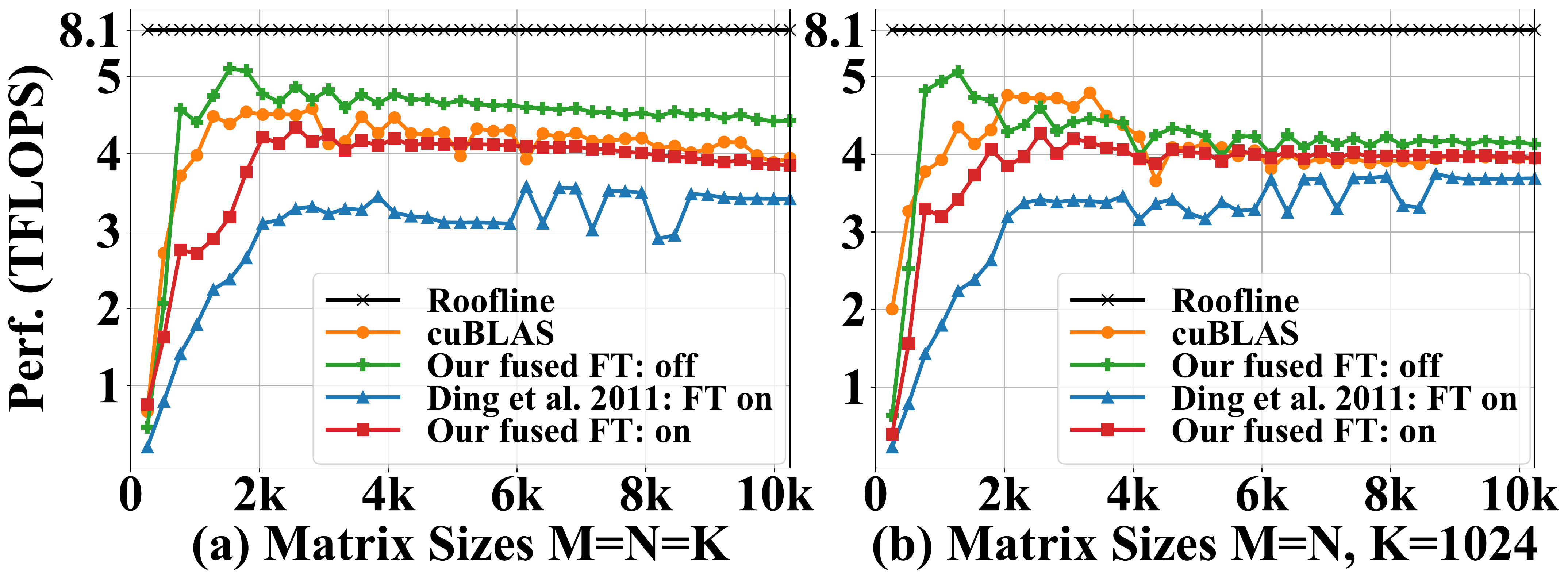} 
    \caption{Comparison of performance with and without fault tolerance on a T4 GPU. CuBLAS and non-fused FT-SGEMM kernel benchmarks are included. The \textit{on} and \textit{off} terms indicate whether the fault tolerance capability has been enabled.}\label{fig:T4_sgemm_vs_ftsgemm}
\end{figure}

Enabling fault-tolerance capability in the fused FT-SGEMM algorithm incurs an average overhead of 11.31\%; however, the algorithm remains comparable in performance to the state-of-the-art cuBLAS library. Figure \ref{fig:T4_sgemm_vs_ftsgemm} compares the performance of cuBLAS SGEMM, non-fused FT-SGEMM, and fused FT-SGEMM with and without fault tolerance. For  $M=N=K$ square matrices, fault tolerance adds an overhead of 14.85\%, while for $M=N, K=1024$, the overhead is 8.55\%. Notably, the fused FT-SGEMM with fault tolerance exhibits a reasonable overhead of 5.33\% to 7.71\% compared to cuBLAS SGEMM. These results demonstrate the effectiveness of the proposed fused ABFT schemes.

\subsubsection{FT-GEMM with codegen}

\begin{figure}[ht]
    \centering
\includegraphics[scale=0.19]{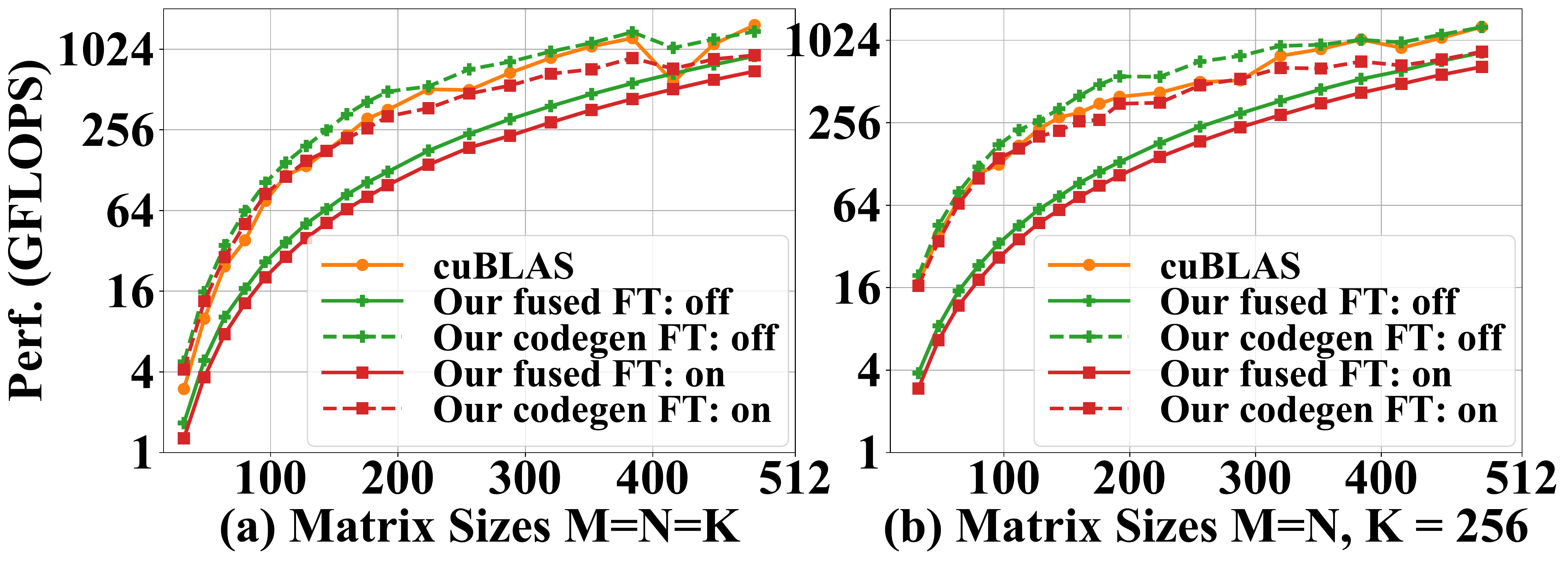}
    \caption{Auto-generated fused FT-SGEMM outperforms original fused FT-SGEMM.}
    \label{fig:T4_why_code_gen}
\end{figure}

The codegen technique further boosts the performance of our FT-SGEMM implementation. Figure \ref{fig:T4_why_code_gen} compares the performance of auto-generated kernels with cuBLAS and the original kernels. The auto-generated FT-SGEMM kernels outperform both the original kernel and cuBLAS SGEMM by 183.50\% and 27.81\% on average, respectively, when the fault tolerance is off. When fault tolerance is on, the improvement of the auto-generated FT-SGEMM kernels is reduced to 165.12\% compared to the original kernel, but the overhead with respect to cuBLAS SGEMM is significantly reduced from 59.23\% to 4.88\%.

\begin{figure}[ht]
    \centering
\includegraphics[scale=0.195]{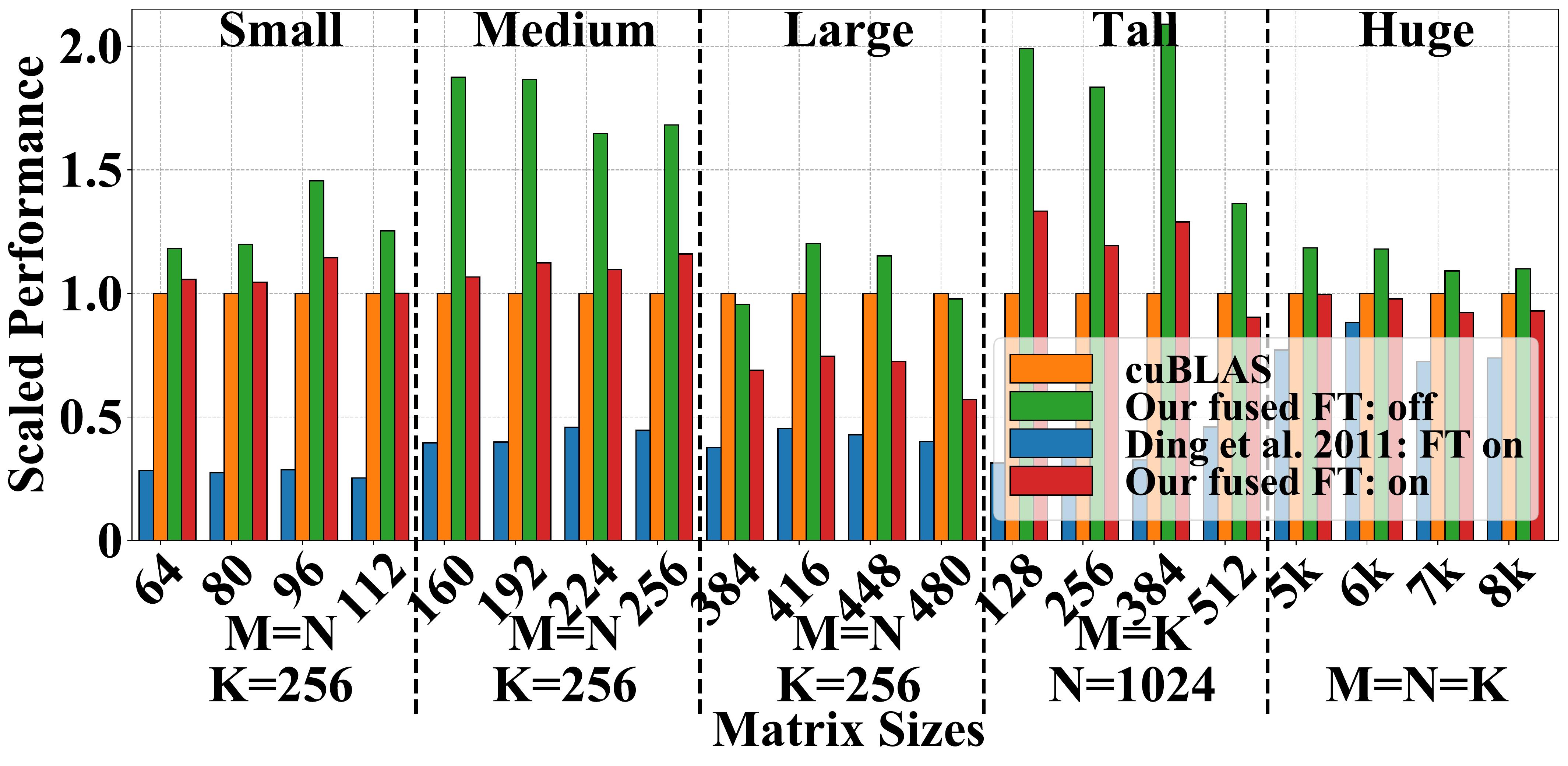} 
    \caption{Performance of generated SGEMM with FT on a Tesla T4 GPU.}
    \label{fig:T4_sgemm_bar}
\end{figure}

Figure \ref{fig:T4_sgemm_bar} compares the performance of five generated fused FT-SGEMM kernels with cuBLAS and non-fused FT-SGEMM for irregular input shapes beyond square inputs. Here we adopt the same templatization, parameterization, and code generation techniques as presented for the baseline SGEMM kernels when generalizing to the fault-tolerant kernels. Our threadblock-level fault-tolerant scheme adds marginal performance overhead to the baseline and maintains a performance comparable to or faster than the closed-source cuBLAS SGEMM. To be more specific, with the fault tolerance functionality turned on, the generated kernel outperforms cuBLAS by 7.22\% to 81.95\% depending on the kernel size, and outperforms the non-fused FT-SGEMM by 64.69\% to 287.06\%.

\subsection{Benchmarking SGEMM with Fault Tolerance under Error Injection}

We validate the effectiveness of our fault-tolerant scheme by injecting multiple computing errors into each of our computing kernels and verify our final computation results against cuBLAS. We choose to inject errors at the source code level to minimize the performance impact on native programs. Since our main scope focuses on compute errors rather than memory errors in the paper, errors are inserted in the register of the accumulated result by adding a numerical offset to emulate register bit flipping. For each thread block, a certain number of errors are evenly injected into random threads throughout the computation. The injected error will lead to a mismatch in the checksum verification step, such that the erroneous element and error magnitude can be computed according to the checksum relationship. The detected error is then corrected by subtracting the error magnitude from the erroneous position. Upon correction of all detected errors, we validate the correctness of our final computation results by comparing them with cuBLAS to ensure the sanity of our fault-tolerant scheme.

\begin{figure}[ht]
    \centering
\includegraphics[scale=0.19]{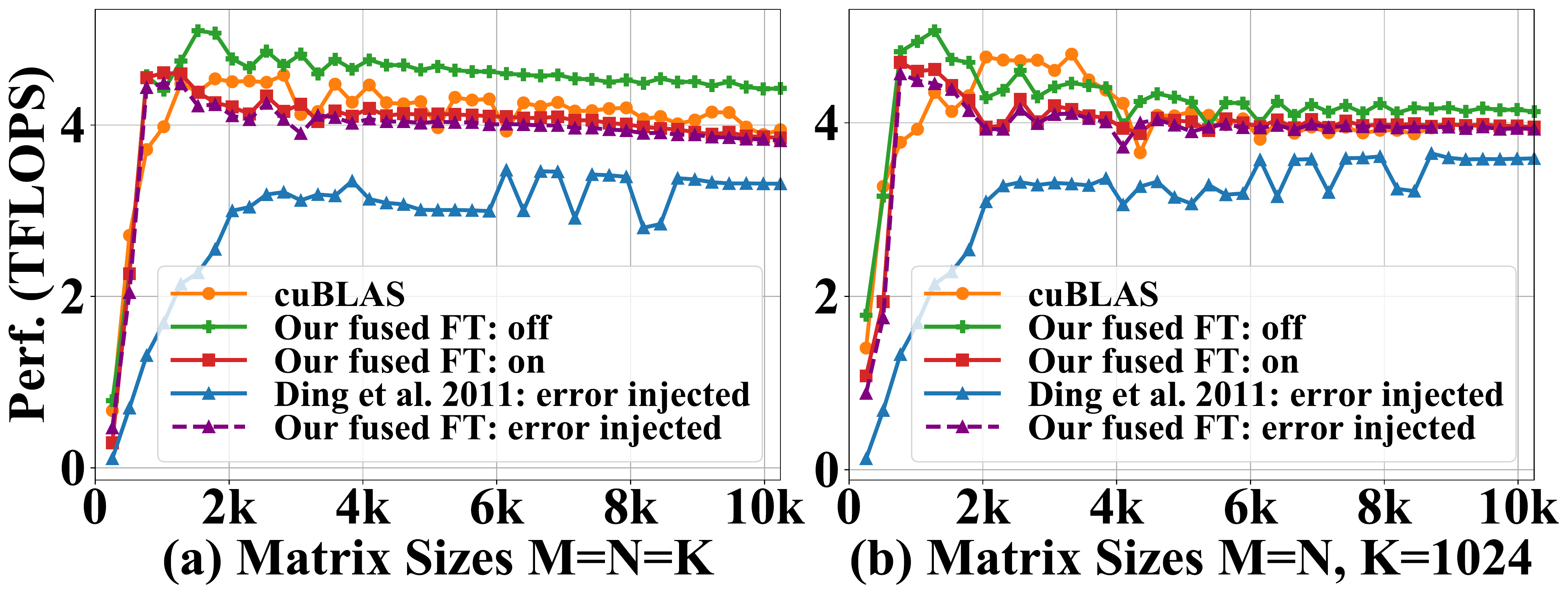} 
    \caption{Error injection on a T4 GPU.}
    \label{fig:T4_error_injection}
\end{figure}

Figure \ref{fig:T4_error_injection} compares the performance of fault-tolerant SGEMM under error injection. Since Ding's ABFT adopts an outer-product-based GEMM, i.e., the output matrix is accumulated over a series of $M$-by-$K_s$ $\times$ $K_s$-by-$N$ GEMMs, errors are injected at each static step size $K_s = 256$ on the $K$ dimension. For a fair comparison, we adopt the same error inject distance $K_s$ as the non-fused ABFT scheme proposed by Ding et al. in 2011 \cite{ding2011matrix}. That is, 1, 2, ..., 40 errors are injected (and corrected) for each outer-product matrix multiplication sub-problem, whose $K$ dimension ranges from $256$ to $10240$. Our experimental results validate that our error correction feature adds minimal extra clock cycles on the detecting-only FT-SGEMM scheme. Compared with the non-fused Ding's ABFT scheme, our fused FT-SGEMM achieves a speedup by $38.8\%$ on average while maintaining a negligible overhead of less than $10\%$. Compared to cuBLAS, our fused FT-SGEMM with error injection has only a $3.22\%\sim4.9\%$ overhead.

\begin{figure}[ht]
    \centering
\includegraphics[scale=0.19]{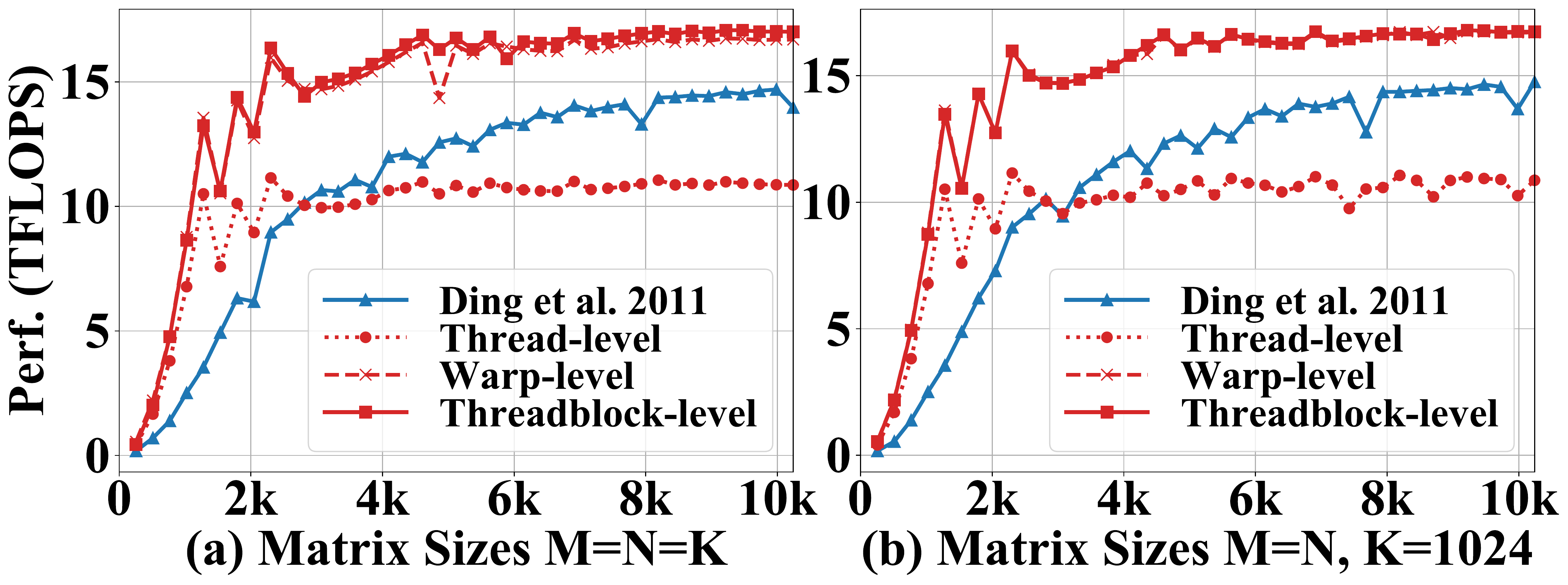} 
    \caption{Different schemes of SGEMM with FT on an A100 GPU.}
    \label{fig:A100_diff_level_ftsgemm}
\end{figure}

\subsection{Performance Evaluation on A100}

In addition to the NVIDIA Tesla T4 GPU, we further evaluate the effectiveness of our optimizations on an NVIDIA A100 GPU. Figure \ref{fig:A100_diff_level_ftsgemm} compares the performance of non-fused FT-SGEMM, thread-level FT-SGEMM, warp-level FT-SGEMM, and threadblock-level FT-SGEMM. The threadblock-level FT-SGEMM outperforms non-fused FT-SGEMM, thread-level FT-SGEMM, and warp-level FT-SGEMM. For the $M=N=K$ case, the threadblock-level FT-SGEMM outperforms the non-fused, thread-level, and warp-level schemes by $52.39\%$, $47.21\%$, and $1.02\%$, respectively. For the $M=N, K=1024$ case, the threadblock-level FT-SGEMM outperforms the non-fused, thread-level, and warp-level schemes by $54.93\%$, $47.18\%$, and $0.04\%$, respectively.

\begin{figure}[ht]
    \centering
\includegraphics[scale=0.19]{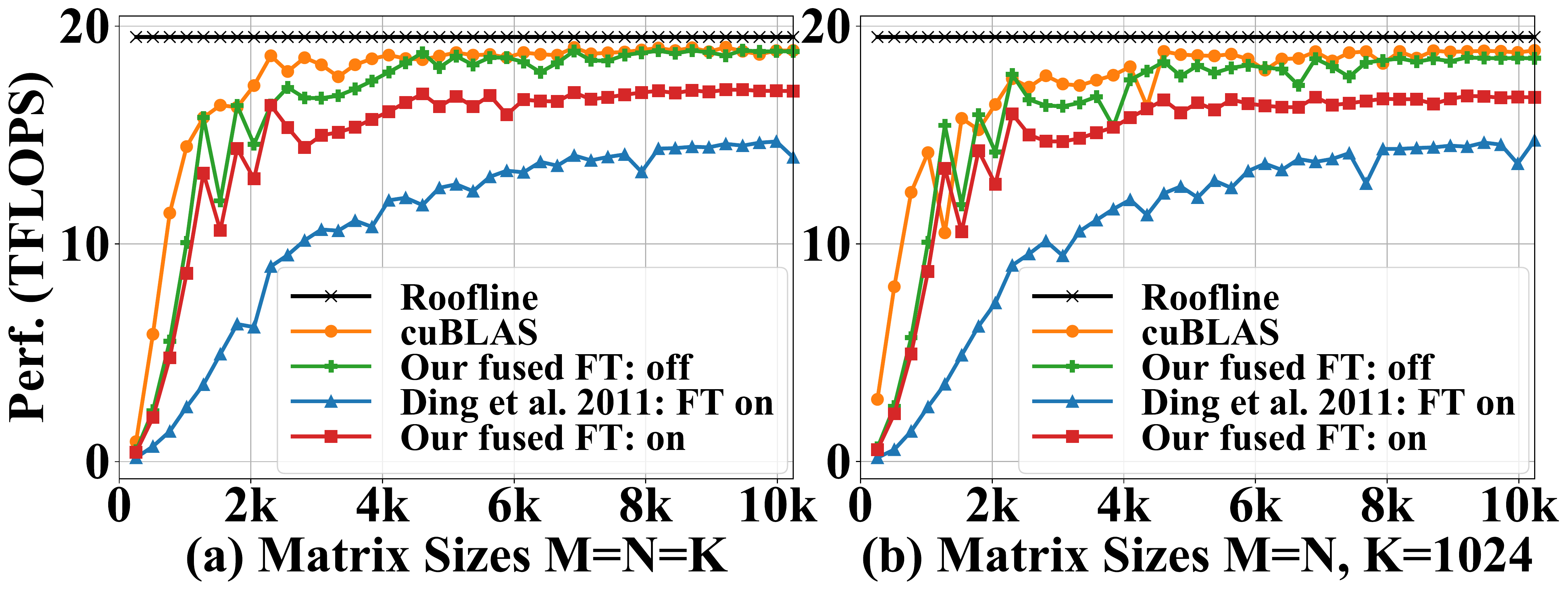} 
    \caption{SGEMM vs. FT-SGEMM on an A100 GPU.}
     \label{fig:A100_sgemm_vs_ftsgemm}
\end{figure}

Figure \ref{fig:A100_sgemm_vs_ftsgemm} compares the performance of cuBLAS SGEMM, baseline ABFT SGEMM, our optimized SGEMM, and our FT-SGEMM. For the $M=N=K$ case, our SGEMM kernel has a $6.29\%$ overhead compared to cuBLAS SGEMM. Our ABFT SGEMM has a $15.32\%$ overhead compared to cuBLAS SGEMM and a $9.93\%$ overhead compared to our SGEMM. For the $M=N, K=1024$ case, our SGEMM kernel has a $3.27\%$ overhead compared to cuBLAS SGEMM. Our ABFT SGEMM has a $15.15\%$ overhead compared to cuBLAS SGEMM, a $9.44\%$ overhead compared to our SGEMM.

\begin{figure}[ht]
    \centering
\includegraphics[scale=0.18]{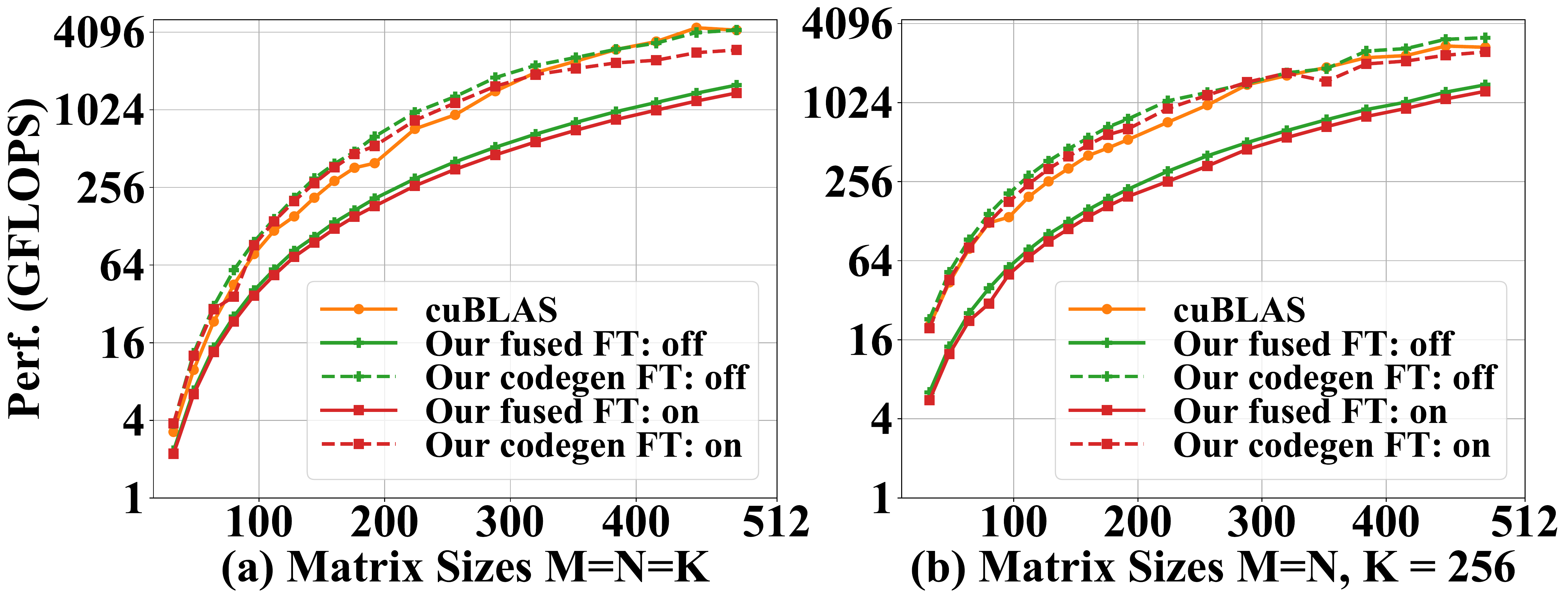} 
    \caption{Code generation on an A100 GPU.}
    \label{fig:A100_why_code_gen}
\end{figure}

Figure \ref{fig:A100_why_code_gen} compares the performance of auto-generated kernels with cuBLAS and original kernels. For the  $M=N=K$ scenario, the auto-generated SGEMM kernels outperform cuBLAS SGEMM by $20.26\%$ and improves our original SGEMM by $160.10\%$. The auto-generated FT-SGEMM kernels outperform cuBLAS SGEMM by $5.94\%$, and the original FT-SGEMM kernel by $148.55\%$. In the case where $M=N, K=256$, the auto-generated SGEMM kernels outperform cuBLAS SGEMM by $22.45\%$ and our original SGEMM by $197.78\%$. The auto-generated FT-SGEMM kernels outperform cuBLAS SGEMM by $7.07\%$, and the original FT-SGEMM kernel by $195.07\%$.

\begin{figure}[ht]
    \centering
\includegraphics[scale=0.19]{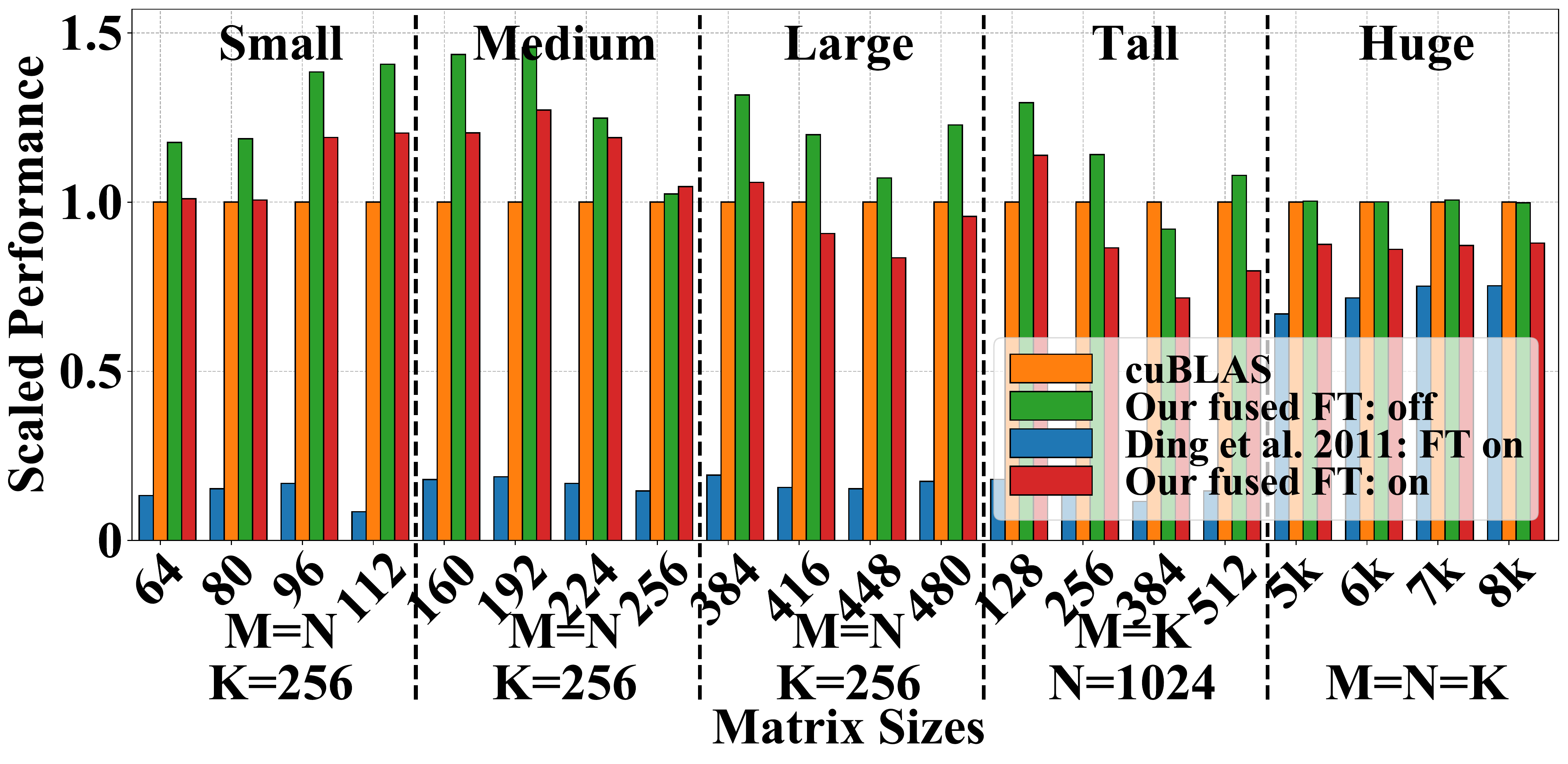} 
    \caption{Performance of generated kernels on an A100 GPU.}
    \label{fig:A100_sgemm_bar}
\end{figure}

Figure \ref{fig:A100_sgemm_bar} compares the performance of generated kernels with cuBLAS and ABFT baseline. Five generated kernels with different sizes are included. For irregularly shaped GEMMs, our heuristic parameter selection mechanism, as shown in Table \ref{tab:kernel_size}, presents a sound improvement over the hard-coded variant. In addition, our fault-tolerant scheme maintains a marginal overhead --- 15.68\% on average over a series of input matrices. Compared with cuBLAS SGEMM, our generated kernels present a slightly faster performance on small inputs and comparable performance for larger matrices. Meanwhile, compared with the non-fused ABFT baseline, our fused ABFT kernels demonstrate a significant improvement, 462.56\% faster on average for small-to-huge input shapes.

\begin{figure}[ht]
    \centering
\includegraphics[scale=0.19]{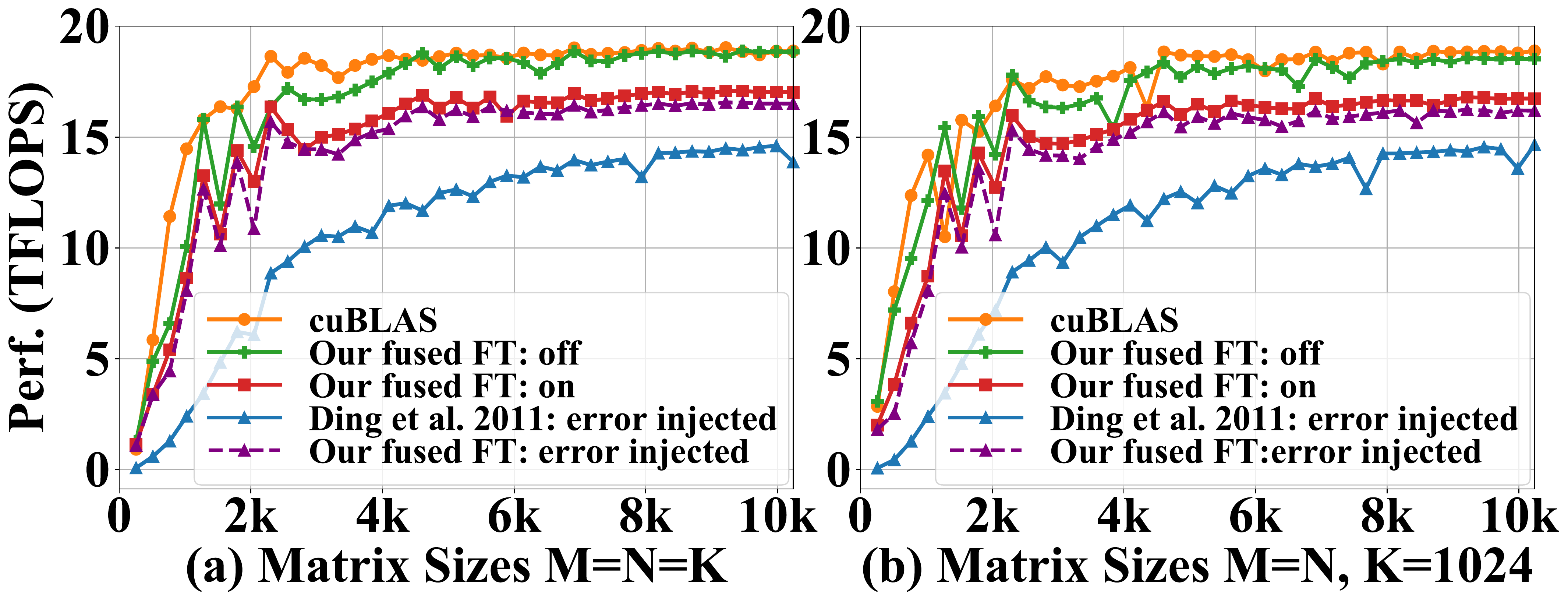} 
    \caption{Error injection experiments on an A100 GPU.}
    \label{fig:A100_error_injection}
\end{figure}

Under tens of error injections, our FT-SGEMM outperforms the non-fused ABFT SGEMM by $56.12\%$. The performance of FT-SGEMM under error injection is $18\%$ slower than the cuBLAS SGEMM without any fault-tolerance capability or error injections, as shown in Figure \ref{fig:A100_error_injection}. For the $M=N=K$ case, our ABFT-SGEMM with error injection is $18.17\%$ slower than cuBLAS SGEMM, a $12.97\%$ overhead compared to our optimized SGEMM, and a $3.39\%$ overhead compared to ABFT-SGEMM without error injection. For the $M=N, K=1024$ case, our ABFT-SGEMM with error injection is $18.40\%$ slower than cuBLAS SGEMM, a $13.03\%$ overhead compared to our optimized SGEMM, and a $3.96\%$ overhead compared to ABFT-SGEMM without error injection.

\subsection{Online ABFT vs. Offline ABFT}

In addition to detailing the optimization techniques of ABFT, we provide insights into our fault-tolerant approaches by comparing them with the detecting-only schemes. Our fault-tolerant approach not only detects errors on-the-fly but corrects these detected errors as well. Therefore, this fault-tolerant scheme is called an \textit{online} ABFT scheme. This online correction mechanism comes with a cost. If only detecting rather than correcting an error, the register budget to track the row and column checksum encoding can be released such that the performance overhead will be significantly alleviated. However, this detecting-only fault-tolerant scheme requires a re-compute when an error is detected, namely offline ABFT, which leads to 100\% overhead in that case.

\begin{figure}[ht]
    \centering
\includegraphics[scale=0.19]{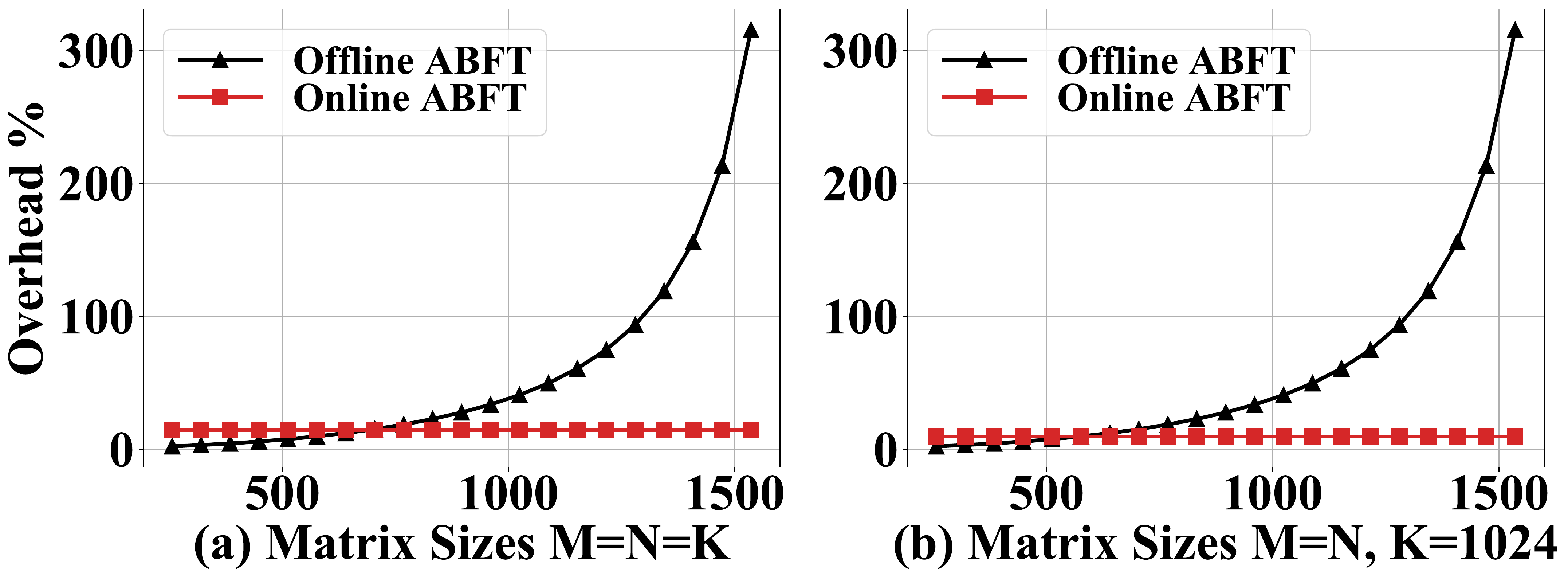} 
    \caption{Online ABFT GEMM vs offline ABFT GEMM.}
    \label{fig:online_vs_offline}
\end{figure}

In practice, whether the online ABFT is more beneficial than the offline version depends on the actual error rate. Consider the following situations: $C+=AB$ with size $M, N, K$. Each threadblock accumulates a tile $C_{tb}$ with size $m_{tb}, n_{tb}$. The accumulation in the threadblock has an error rate of $\gamma_0 \in (0,1)$, namely each accumulation has a probability of $\gamma_0$ for an error occurence. Hence, the overall error rate is $\gamma = 1-(1-\gamma_0)^{\frac{M}{m_{tb}}\times \frac{N}{n_{tb}}}$. Since the offline ABFT requires a restart for errors that occur during the recomputation, the expected restart times to obtain a correct SGEMM result is $(1-\gamma) + 2\gamma ((1-\gamma) + 2\gamma(\cdots)) = (1-\gamma)(1 + 2\gamma  + (2\gamma)^2 + \cdots) = \frac{(1-\gamma)}{1-2\gamma}$. In contrast, the online ABFT scheme provides a precise error correction on-the-fly so the expected computation times to obtain a correct result is just $1$. For the error model presented above, we compare the overhead of online FT-SGEMM and offline FT-SGEMM under an error rate $\gamma_0 = \frac{1}{256}$ in Figure \ref{fig:online_vs_offline}. Following the idea presented in \cite{kosaian2021arithmetic}, we implement an offline ABFT for SGEMM that only detects errors on-the-fly. Due to its significantly smaller register usage at runtime for fault tolerance, we observe a performance overhead close to 1\%.

\section{Conclusion}
\label{sec:conclusion}

In this study, we proposed a high-performance GEMM design that is equipped with algorithm-based fault tolerance specifically for GPUs. Our design includes fault-tolerant designs for GEMM at the thread, warp, and threadblock levels, and a baseline GEMM implementation that rivals or exceeds the performance of the state-of-the-art closed-source cuBLAS GEMM. Moreover, we presented a cost-efficient template-based approach for code generation that supports a wide range of input matrix shapes. Our future works will concentrate on broadening the scope of our strategies to include additional data types, such as ZGEMM for double-precision complex matrices, as well as for GPUs from AMD and Intel.

\section{Acknowledgement}

This work was supported by the U.S. Department of Energy, Office of Science, Office of Advanced Scientific Computing Research, Scientific Discovery through the Advanced Computing (SciDAC) program under Award Number DE-SC0022209. We thank the anonymous reviewers for their insightful comments.

\bibliographystyle{ACM-Reference-Format}
\bibliography{sample-base}

\end{document}